\documentclass[aps,pra,reprint,superscriptaddress,showpacs]{revtex4-1}
\usepackage{graphicx}
\usepackage{amsmath,amsfonts,bm}
\usepackage{color}

\begin{document}
% Use the \preprint command to place your local institutional report
% number in the upper righthand corner of the title page in preprint mode.
% Multiple \preprint commands are allowed.
% Use the 'preprintnumbers' class option to override journal defaults
% to display numbers if necessary
%\preprint{}

%Title of paper
\title{Full-wave analytical solution of second-harmonic generation in metal nanospheres}

\author{Antonio Capretti}
\affiliation{Department of Electrical Engineering and Information Technology, Universit\`{a} degli Studi di Napoli Federico II, Via Claudio 21, 80125 Napoli, Italy}
\affiliation{Department of Electrical and Computer Engineering \& Photonics Center, Boston University, 8 Saint Mary’s Street Boston, MA 02215, United States}
\author{Carlo Forestiere}
\affiliation{Department of Electrical Engineering and Information Technology, Universit\`{a} degli Studi di Napoli Federico II, Via Claudio 21, 80125 Napoli, Italy}
\affiliation{Department of Electrical and Computer Engineering \& Photonics Center, Boston University, 8 Saint Mary’s Street Boston, MA 02215, United States}
\author{Luca Dal Negro}
\affiliation{Department of Electrical and Computer Engineering \& Photonics Center, Boston University, 8 Saint Mary’s Street Boston, MA 02215, United States}
\author{Giovanni Miano}
\email[Corresponding author: ]{miano@unina.it}
\affiliation{Department of Electrical Engineering and Information Technology, Universit\`{a} degli Studi di Napoli Federico II, Via Claudio 21, 80125 Napoli, Italy}

\date{\today}

\begin{abstract}
We present a full-wave analytical solution for the problem of second-harmonic generation from spherical nanoparticles.
The sources of the second-harmonic radiation are represented through an effective nonlinear polarization.
The solution is derived in the framework of the Mie theory by expanding the pump field, the nonlinear sources and the second-harmonic fields in series of spherical vector wave functions.
We use the proposed solution for studying the second-harmonic radiation generated from gold nanospheres as function of the pump wavelength and the particle size, in the framework of the Rudnick-Stern model.
We demonstrate the importance of high-order multipolar contributions to the second-harmonic radiated power.
Moreover, we investigate the p- and s- components of the SH radiation as the Rudnick-Stern parameters change, finding a strong variation.
This approach provides a rigorous methodology to understand second-order optical processes in metal nanoparticles, and to design novel nanoplasmonic devices in the nonlinear regime.
\end{abstract}

% insert suggested keywords - APS authors don't need to do this
%\keywords{}

%\maketitle must follow title, authors, abstract, \pacs, and \keywords
\maketitle

\section{Introduction}
Nonlinear phenomena in metal nanostructures are gathering much attention due to their potential application as novel components for integrated optics \cite{Maier2012,Brevet2010NL,Capretti2012,Panoiu2010,Kauranen2012}. Moreover, second-harmonic (SH) generation from metal nanostructures provides a powerful tool for probing physical and chemical properties of material surfaces \cite{Cheng2011,Gonella2011}.

Noble metal nanoparticles support Localized Surface Plasmons (LSPs). LSPs are collective oscillations of the conduction electrons, that strongly affect the optical response of the metal. When LSPs are resonantly excited, the local electromagnetic field is significantly enhanced in the particle, enabling nonlinear optical effects, such as harmonic generation, at relatively low excitation powers.

Second-harmonic radiation originates from two contributions: the particle bulk and the surface, respectively. In noble metal nanoparticles, the local-bulk source is absent because of the material centrosymmetry, and only the nonlocal-bulk contribution needs to be considered \cite{Heinz}. The surface contribution to SH radiation is due to the symmetry breaking at the interface with the embedding medium \cite{Agarwal1982,Heinz1999}. The magnitudes of the nonlocal-bulk and surface SH contributions depend on the shape of the nanoparticle and on the optical properties of the metal at the fundamental and second-harmonic frequencies \cite{Shen1986,Shen1987,Sibilia2011,Smith2012,Scalora2010,Moloney2009,Kujala2007,Husu2012}.

In 1999, Dadap \textit{et al.} studied the SH radiation generated from the surface of a sphere with radius R much smaller than the wavelength of the incident light $\lambda$ ($2 \pi R / {\lambda} << 1$, Rayleigh limit), made of a centrosymmetric and isotropic material \cite{Heinz1999}.
They showed that the leading-order contributions to SH radiation arise from the electric-dipole ${{\bf{p}}^{\left( {2\omega } \right)}}$ and the electric-quadrupole ${{\tensor{\bf Q}^{(2\omega )}}}$ moments, and gave the main selection rules for the SH scattering from a sphere.
In Refs. \onlinecite{Brudny2000} and \onlinecite{Heinz2004}, the Rayleigh limit is analyzed by taking into account both the bulk and the surface polarization sources, showing that the SH field is radiated by an effective electric dipole moment
${\bf{p}}_{eff}^{(2\omega )} ({\bf{\hat r}})\cong {{\bf{p}}^{(2\omega )}} + i~{k_0}{{\tensor{\bf Q}^{(2\omega )}}}{\bf{\hat r}} / 3$
(the SH magnetic dipole emission is forbidden because of the axial symmetry of the system).
In the Rayleigh limit, both the nonlocal-bulk and the surface SH sources contribute to the induced electric dipole moment, while only the local surface sources contribute to the induced electric quadrupole moment.
The presence of distinct SH sources with their own radiation patterns causes the SH Rayleigh scattering process to differ significantly from the linear Rayleigh scattering. In particular, the theory predicts the absence of the SH signal in the forward direction and the ${\left( {{2\pi}R / \lambda} \right)^6}$ scaling of the SH scattering cross-section.

The SH Rayleigh scattering model is inaccurate if the particle size is comparable with the wavelength, because the contributions of SH multipolar orders higher than 2 are not negligible. A full-wave analysis of the SH scattering from spherical particles of arbitrary size is developed in Ref. \onlinecite{Pavlyukh2004}, but only the surface SH source was taken into account. Moreover, the enforced boundary conditions are incorrect \cite{Heinz}, resulting in zero SH radiation from the radial component of the source. Recently, a full wave theory of the SH radiation generated by a chain of parallel infinitely long cylinders, including both the bulk and surface nonlinear sources, has been developed in Ref. \onlinecite{Panoiu2010}. A full-wave theory of the SH radiation generated in three-dimensional structures consisting of metal spheres made of centrosymmetric materials has been proposed in Ref. \onlinecite{Zhang2012}, but here again the treatment is limited to the surface source.

The SH sources in metal nanoparticles can be represented by an effective nonlinear polarization induced by the electromagnetic field at the fundamental frequency $\omega$.
As noble metals are isotropic and centrosymmetric materials, the bulk contribution ${\bf{P}}_b^{^{\left( {2\omega } \right)}}$ to the nonlinear polarization is of the form \cite{Bloembergen1968}:
\begin{equation}
\label{Pbulk}
\begin{split}
{\bf{P}}_b^{^{\left( {2\omega } \right)}} =
& {\varepsilon _0} \; \beta {{\bf{E}}^{\left( \omega  \right)}}\nabla  \cdot {{\bf{E}}^{\left( \omega  \right)}} +
  {\varepsilon _0} \; \gamma \nabla \left( {{{\bf{E}}^{\left( \omega  \right)}} \cdot {{\bf{E}}^{\left( \omega  \right)}}} \right) +\\
& {\varepsilon _0} \; \delta' \left( {{{\bf{E}}^{\left( \omega  \right)}} \cdot \nabla } \right){{\bf{E}}^{\left( \omega  \right)}}
\text{~~~~~~~~~~~~~~~~~~in } \Omega_i \text{ ,}
\end{split}
\end{equation}
where $\beta$, $\gamma$ and $\delta'$ are material parameters, $\varepsilon_0$ is the vacuum permittivity, $\bm{E}^{(\omega)}$ is the electric field at the fundamental frequency and $\Omega_i$ denotes the region occupied by the particle.
Due to the homogeneity of the material we also have $\nabla\cdot\bm{E}^{(\omega)}=0$ in $\Omega_i$, therefore the first term on the right hand side of Eq. (\ref{Pbulk}) vanishes and the expression of ${\bf{P}}_b^{\left( {2\omega } \right)}$ reduces to:
\begin{equation}
%\label{Pbulk}
{\bf{P}}_b^{^{\left( {2\omega } \right)}} =
  {\varepsilon _0} \; \gamma \nabla \left( {{{\bf{E}}^{\left( \omega  \right)}} \cdot {{\bf{E}}^{\left( \omega  \right)}}} \right) +
  {\varepsilon _0} \; \delta' \left( {{{\bf{E}}^{\left( \omega  \right)}} \cdot \nabla } \right){{\bf{E}}^{\left( \omega  \right)}}
\text{.}
\end{equation}
The surface contribution ${\bf{P}}_s^{^{\left( {2\omega } \right)}}$ is of the form \cite{Heinz}:
\begin{equation}\label{eq:Psurf1}
{\bf{P}}_s^{\left( {2\omega } \right)} =
{\varepsilon _0} \;
\mathord{\buildrel{\lower3pt\hbox{$\scriptscriptstyle\leftrightarrow$}} \over \chi } _s^{\left( {2\omega } \right)}{\rm{:}}
\left. {{\bf{E}}^{\left( \omega  \right)}}{{\bf{E}}^{\left( \omega  \right)}} \right| _{\Sigma_i}
\text{ on } \Sigma \text{ , }
\end{equation}
where ${\mathord{\buildrel{\lower3pt\hbox{$\scriptscriptstyle\leftrightarrow$}} \over \chi } _s^{\left( {2\omega } \right)}}$ is the second-order surface nonlinear susceptibility tensor of the metal, and $\Sigma$ denotes the particle boundary.
The normal component of $\bm{E}^{(\omega)}$ is evaluated on the internal page of $\Sigma$, which we have indicated with $\Sigma_i$;
there is no ambiguity relevant to the tangential components of $\bm{E}^{(\omega)}$ because they are continuous across $\Sigma$.
Since the nanoparticle surface possesses isotropic symmetry with a mirror plane perpendicular to it, the tensor ${\mathord{\buildrel{\lower3pt\hbox{$\scriptscriptstyle\leftrightarrow$}} \over \chi } _s^{\left( {2\omega } \right)}}$ has only three non-vanishing and independent elements, ${{\chi _{ \bot  \bot  \bot }}}$, ${{\chi _{ \bot \parallel \parallel }}}$ and ${{\chi _{\parallel  \bot \parallel }} = {\chi _{\parallel \parallel  \bot }}}$, where $\bot$ and $\parallel$ refer to the orthogonal and tangential components to the particle surface \cite{Heinz}.
% (Fig. \ref{figure01}(a))
Therefore Eq. (\ref{eq:Psurf1}) reduces to:
\begin{equation}
\begin{split}
{\bf{P}}_s^{\left( {2\omega } \right)} \cong {\varepsilon _0}
\left[ {{ \chi _{ \bot  \bot  \bot }}{\bf{\hat n\hat n\hat n}} + }\right.
\left. {{ \chi _{\bot \parallel \parallel }}\left( { {\bf{\hat n}}{{{\bf{\hat t}}}_1}{{{\bf{\hat t}}}_1} + {\bf{\hat n}}{{{\bf{\hat t}}}_2}{{{\bf{\hat t}}}_2}} \right)} \right. + \\
\left. {{ \chi _{\parallel  \bot \parallel }}\left( {{{{\bf{\hat t}}}_1}{\bf{\hat n}}{{{\bf{\hat t}}}_1} + {{{\bf{\hat t}}}_2}{\bf{\hat n}}{{{\bf{\hat t}}}_2}} \right)} \right]
{\rm{:}}{{\bf{E}}^{\left( \omega  \right)}}{{\bf{E}}^{\left( \omega  \right)}} \text{ , }
\end{split}
\end{equation}
where ${\bf{\hat n}}$ is the normal to the particle surface pointing outward and ${\bf{\hat t}_1}$,${\bf{\hat t}_2}$ are two orthonormal vectors defining the plane tangent to the particle surface, such that $({\bf{\hat n}},{\bf{\hat t}_1}$,${\bf{\hat t}_2})$ is a counterclockwise triplet.
It is interesting to note that, although the the relation between ${\bf{P}}_s^{\left( {2\omega } \right)}$ and ${{\bf{E}}^{\left( \omega  \right)}}$ is of local character, the contribution of the normal component $({\bf{P}}_s^{\left( {2\omega } \right)} \cdot {\bf{\hat n}})$ to the SH radiation depends on the surface gradient ${\nabla _S} ( {{\bf{P}}_s^{\left( {2\omega } \right)} \cdot {\bf{\hat n}}} )$, as shown in Appendix \ref{app:selvedge}.

The theoretical and experimental determination of the parameters $\gamma$, $\delta'$, $\chi _{ \bot  \bot  \bot }$, $\chi _{ \bot  \parallel  \parallel }$ and $\chi _{ \bot  \parallel  \bot }$ has been a long-standing problem in Nonlinear Optics, and it is still an open issue \cite{Stegeman1980,Stegeman1987,Kauranen2009,Brevet2010PRB}.
% could be obtained by modeling the basic bulk and surface mechanisms that originate the SH radiation, or by identifying them through appropriate measurement schemes, or by combining both theoretical and experimental identifications.
The source of the nonlinearity in metals results from the response of both bound and free electrons. In particular, for the visible/near-IR part of the light spectrum, the nonlinear response of thick metal particles may be attributed mostly to the free electrons \cite{Bloembergen1968,Stegeman1980,Scalora2010,Smith2012}.
They behave as an isotropic electron gas with effective mass $m_{eff}$, relaxation time $\tau$ (due to the collisions with the ion lattice) and a quantum pressure.
The electron gas dynamics are governed by the Euler's equation. This is the so-called \textit{hydrodynamic model}.
Within it, the bulk contribution to the nonlinear polarization arises from both the convective term and the Lorentz's force term, while the surface contributions are strictly related to the response of the electrons within the Thomas-Fermi screening length ($\lambda_{TF}\approx 1\AA$ for gold) from the surface \cite{Stegeman1980,Smith2012PRX}. Since in our case $\omega\tau>>1$, the hydrodynamic model gives the following expressions for the bulk parameters $\gamma$ and $\delta'$ \cite{Bertolotti2009,Kauranen2009}:
\begin{subequations}
\begin{equation}
\gamma  =  - \frac{1}{8} {\chi_b}\left( \omega  \right)\frac{{\omega _p^2}}{{{\omega ^2}}}\frac{{\varepsilon _0}}{{e{n_0}}}
\end{equation}
\begin{equation}
\delta' \cong i \frac{2 \gamma}{\omega\tau}
\end{equation}
\end{subequations}
where $\chi_b(\omega)=\varepsilon_i (\omega) / \varepsilon_0 -1$ is the linear bulk permittivity of the metal, $n_0$ is the equilibrium number density of the free electrons, $\omega_p=\sqrt{n_0 e^2 / m_{eff} \varepsilon_0}$ is the free electron plasma frequency  and $-e$ is the electron charge.
In the same limit $\omega\tau>>1$, the hydrodynamic model also gives the following estimation for the surface parameters $\chi _{ \bot  \bot  \bot }$ and $\chi _{\parallel \bot  \parallel }$ \cite{Stegeman1980,Stegeman1987}:
\begin{subequations}
\begin{align}
{\chi _{ \bot  \bot  \bot }} =  - \frac{1}{4} &{\chi_b}\left( \omega  \right)\frac{{\omega _p^2}}{{{\omega ^2}}}\frac{{\varepsilon _0}}{{e{n_0}}}, \\
{\chi _{\parallel  \bot \parallel }} = \frac{1}{2} &{\chi_b}\left( \omega  \right)\frac{{\omega _p^2}}{{{\omega ^2}}}\frac{{\varepsilon _0}}{{e{n_0}}},
\end{align}
\end{subequations}
Furthermore, the contribution of the term $\chi _{ \bot  \parallel  \parallel }$ is negligible \cite{Stern1971,Stegeman1980,Kauranen2009,Brevet2010PRB}.

Alternatively, the parameters of the SH sources may be identified experimentally. This would allow to account for phenomena that are disregarded in the hydrodynamic model, as the interband transitions.
Nevertheless, an identification procedure of the parameters $\gamma$, $\delta'$, ${\chi _{ \bot  \bot  \bot }}$, ${\chi _{ \bot  \parallel  \parallel}}$ and ${\chi _{\parallel \bot \parallel}}$ through measurements of the SH radiation has an intrinsic limit.
In fact, the parameter $\gamma$ cannot be separated from the surface terms ${\chi _{ \bot  \bot  \bot }}$ and ${\chi _{ \bot  \parallel  \parallel}}$, through measurements of the SH field outside the metal.
An equivalent surface nonlinear polarization with surface susceptibility $\chi_{eff}\left[ \bm{\hat n}\bm{\hat n}\bm{\hat n} +\bm{\hat n}(\bm{\hat t}_1 \bm{\hat t}_1 + \bm{\hat t}_2 \bm{\hat t}_2 ) \right]$, where $\chi_{eff}=\gamma(\omega) \varepsilon_0 / \varepsilon_i(2\omega)$, generates outside the metal the same electromagnetic field generated by the $\gamma$ term.
For this reason the contribution of the $\gamma$ term is called \textit{surface-like} bulk term \cite{Kauranen2009}.
This is an ancient problem in Nonlinear Optics.
Sipe \textit{et al.} pointed out this ambiguity for the first time by analyzing the SH radiation generated by a planar slab \cite{Stegeman1987}.
Moreover, they inferred that this property holds true for any material and shape.
In this paper we also provide a very simple demonstration of this general property.
On the contrary, Wang \textit{et al.} unambiguously determined the $\delta'$ bulk term in the SH radiation generated from a gold film, by using a two-beam SH generation measurement technique \cite{Kauranen2009}.
However, the Authors conclude "\dots that the surface-like contributions dominate and that the pure bulk component makes only a minor contribution" to the SH radiation generated from a planar slab.
Since the contribution of the $\delta'$ term depends on the spatial derivatives of ${{\bf{E}}^{\left( \omega  \right)}}$, its magnitude may be significant when ${{\bf{E}}^{\left( \omega  \right)}}$ is rapidly varying in the bulk of the metal \cite{Kauranen2009}.
The importance of the $\delta'$ contribution to the SH radiation from non-planar geometries is still an open problem.

In this paper, we propose a full-wave analytical solution for the SH scattering from nanospheres of arbitrary size.
We use this solution to investigate the SH radiation generated by a gold metal nanosphere as function of the polarization, the pump wavelength and the particle size.
We rigorously investigate the multipolar nature of the SH generation and the contributions of the different sources of second-order nonlinearity.
Following Ref. \onlinecite{Brevet2010PRB}, we adopt the Rudnick-Stern model to represent the SH sources of the gold nanosphere \cite{Stern1971}.
In this model the contribution arising from the $\delta'$ term is considered negligible.
This approximation is valid if the field at the fundamental frequency inside the nanoparticle is not rapidly varying, \textit{i.e.} the intensities of high-order multipoles are negligible.
We compare the results obtained by using, as Rudnick-Stern parameters, the Sipe's model values (Ref. \onlinecite{Stegeman1980}) and the values identified experimentally in Ref. \onlinecite{Brevet2010PRB}.
As the Rudnick-Stern parameters vary, we find strong changes of the SH p- and s-components.
%In this paper we propose a full-wave analytical solution for the SH scattering from nanospheres of arbitrary size, which includes both the contributions of the nonlocal-bulk and the local-surface SH sources. As a case of considerable interest, we apply this solution to investigate the SH radiation generated by noble metal nanospheres.

The present work is organized as follows. In Section \ref{formulation}, the electromagnetic formulation of SH scattering by a metal nanosphere is presented, and the analytical solution for the fields is defined, both at the pump and at the SH frequencies. In Section \ref{results}, the SH scattering from gold nanospheres with increasing size is studied, as function of the wavelength and polarization of the pump field. In Section \ref{concl}, the conclusions are outlined. This manuscript is completed by six Appendices, where detailed formulas for the analytical calculation of all the quantities of interest are provided.

\section{Problem statement and solution}\label{formulation}
\subsection{Problem statement}
\begin{figure}
\includegraphics{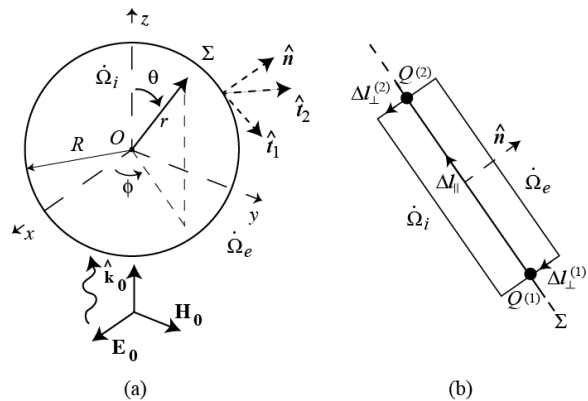}%
\caption{\label{figure01}(a) Scheme of the spherical particle and coordinate system $\left( {O,r,\theta ,\phi } \right)$. (b) Elementary closed curve ${\Delta}l$ across the \textit{selvedge region} at the particle boundary.}
\end{figure}

Let us consider the electromagnetic field at frequency $2\omega$ generated from a metal sphere of radius $R$, when illuminated by a time-harmonic electromagnetic plane-wave at frequency $\omega$ incoming from infinity. We use a spherical coordinate system $\left( {O,r,\theta ,\phi } \right)$ with the origin $O$ in the center of the sphere, as in Fig. \ref{figure01}(a); we denote with
$\left( { \bm{\hat r}, \bm{\hat \theta}, \bm{\hat \phi} } \right)$
the unit vectors of the spherical coordinate system. The domain of the electromagnetic field is the entire space $\mathbb{R}^3$, divided into the interior part of the metal domain ${\dot \Omega _i}$, the embedding medium ${\dot \Omega _e}$ and the metal surface $\Sigma $. The surface $\Sigma $ is oriented in such a way that its normal ${\bf{\hat n}}$  points outward, ${\bf{\hat n}} = {\bf{\hat r}}\left| {_\Sigma } \right.$. We use the convention
${\bf{a}}\left( {{\bf{r}},t} \right) = Re\left\{ {{{\bf{A}}^{\left( \Omega  \right)}}\left( {\bf{r}} \right)\exp \left( {i\Omega t} \right)} \right\}$
for representing a time harmonic electromagnetic field at angular frequency $\Omega$, where ${\bf{r}} = r{\bf{\hat r}}$.

The second-harmonic generation problem involves two electromagnetic fields oscillating at different frequencies:
the electromagnetic field
%\textit{the fundamental-frequency} electromagnetic field
$\left( {{{\bf{E}}^{\left( \omega  \right)}},{\rm{ }}{{\bf{H}}^{\left( \omega  \right)}}} \right)$
at fundamental frequency $\omega $ and the \textit{second-harmonic} electromagnetic field
$\left( {{{\bf{E}}^{\left( {2\omega } \right)}},{\rm{ }}{{\bf{H}}^{\left( {2\omega } \right)}}} \right)$
at frequency $2\omega $. We denote with $\left( {{\bf{E}}_0^{\left( \omega  \right)},{\rm{ }}{\bf{H}}_0^{\left( \omega  \right)}} \right)$
the \textit{incident (pump)} electromagnetic field:
\begin{equation}
\begin{split}
{\bf{E}}_0^{\left( \omega  \right)} &= {E_0} {\bm {\hat \varepsilon} _0}{e^{ - i{\bf{k}}_0^{\left( \omega  \right)} \cdot {\bf{r}}}}\\
{\bf{H}}_0^{\left( \omega  \right)} &= {\frac{{{E_0}}}{{{\zeta _e }}}} \left( {{{{\bf{\hat k}}}_0} \times {{\bm {\hat \varepsilon} }_0}} \right){e^{ - i{\bf{k}}_0^{\left( \omega  \right)} \cdot {\bf{r}}}} \text{ , }
\end{split}
\end{equation}
where ${E_0}$ is the electric field amplitude of the linearly-polarized pump beam, ${\bm{\hat \varepsilon _0}}$ is its polarization direction, ${{\bf{\hat k}}_0}$ is its propagation direction, ${\bf{k}}_0^{(\omega )} = {{\bf{\hat k}}_0}{k_e(\omega )}$,
${k_e(\omega )} = \omega \sqrt {{\mu _e}{\varepsilon _e}}$ and
${\zeta _e} = \sqrt {{\mu _e}/{\varepsilon _e}}$.
%}
The parameters ${\varepsilon _e}$ and ${\mu _e}$ are the permittivity and the permeability of the embedding medium.

Since the intensities of the SH fields generated by noble metals are always orders of magnitude weaker than the intensities of the pump fields, the SH fields do not significantly couple back to the fundamental fields (\textit{undepleted-pump approximation}). As a result, the electromagnetic scattering problems at the fundamental frequency and at the SH frequency are both linear. The linear electromagnetic response of the metal is characterized, in the frequency domain, by the permittivity ${\varepsilon _i}$, which depends on the frequency, and by the permeability ${\mu _i}$, that we assume independent of the frequency.

In order to calculate the SH radiation generated by the metal sphere, we have to evaluate:
\begin{enumerate}
\item the electric field ${{\bf{E}}^{\left( \omega  \right)}}$ in ${\dot \Omega _i}$ and on the inner page of $\Sigma$ (that we have denoted with $\Sigma_i$), induced by the pump electromagnetic field $\left( {{\bf{E}}_0^{\left( \omega  \right)},{\rm{ }}{\bf{H}}_0^{\left( \omega  \right)}} \right)$;
\item the SH nonlinear polarization sources generated by ${{\bf{E}}^{\left( \omega  \right)}}$;
\item the electromagnetic fields $\left( {{{\bf{E}}^{\left( {2\omega } \right)}},{\rm{ }}{{\bf{H}}^{\left( {2\omega } \right)}}} \right)$ radiated by the SH nonlinear polarization fields.
\end{enumerate}
Both problems 1) and 3) are solved by expressing the unknown fields in terms of Spherical Vector Wave Functions (SVWFs), defined in Section \ref{PumpField}.
%
%The order of magnitude of the bulk SH polarization is thus given by:
%\begin{equation}
%P_b^{\left( {2\omega } \right)} = {\varepsilon _0}
%\left[ {{{\left( {\frac{{{\lambda}}}{{{\lambda _p}}}} \right)}^2}\frac{{{E_0}}}{{{E_b}}}} \right]{\chi _b}\left( \omega  \right){E_0} \text{ , }
%\end{equation}
%where ${\lambda _p} = {c_0}\left( {2\pi /{\omega _p}} \right)$ and ${E_b} = \left( {{\lambda _p}/2\pi } \right)e{n_0}/{\varepsilon _0}$. For noble metal nanoparticles ${\lambda _p} \cong 140~nm$ and ${E_b} \approx {10^{14}}{\rm{ V/m}}$. The order of magnitude of the surface SH source is instead given by
%$P_s^{\left( {2\omega } \right)} =
% P_b^{\left( {2\omega } \right)} \left( {{\lambda _p}/2\pi } \right)$.
%
%It is worth noting that the proposed analytical solution of the SH generation from a sphere is independent of the expression of
%$\chi _{ \bot  \bot  \bot }$, 
%$\chi _{\parallel  \bot \parallel }$ and
%$\gamma$.

\subsection{\label{PumpField}Electromagnetic fields at pump frequency}
The electromagnetic field at the fundamental frequency is the solution of the Maxwell's equations:
\begin{subequations}
\label{eq:MaxFF}
\begin{equation}
\label{eq:MaxFF_int}
\begin{cases}
{\nabla  \times {\bf{E}}_i^{\left( \omega  \right)} =  - i\omega {\mu _i}{\bf{H}}_i^{\left( \omega  \right)}}\\
{\nabla  \times {\bf{H}}_i^{\left( \omega  \right)} =  + i\omega {\varepsilon _i}\left( \omega  \right){\bf{E}}_i^{\left( \omega  \right)}}
\end{cases} \text{in } \Omega_i \text{ , }
\end{equation}
\begin{equation}
\label{eq:MaxFF_bc}
\begin{cases}
{{\bf{n}} \times \left( {{\bf{E}}_i^{\left( \omega  \right)} - {\bf{E}}_{sc}^{\left( \omega  \right)}} \right) = {\bf{n}} \times {\bf{E}}_0^{\left( \omega  \right)}}\\
{{\bf{n}} \times \left( {{\bf{H}}_i^{\left( \omega  \right)} - {\bf{H}}_{sc}^{\left( \omega  \right)}} \right) = {\bf{n}} \times {\bf{H}}_0^{\left( \omega  \right)}}
\end{cases} \text{on } \Sigma \text{ , }
\end{equation}
\begin{equation}
\label{eq:MaxFF_ext}
\begin{cases}
{\nabla  \times {\bf{E}}_{sc}^{\left( \omega  \right)} =  - i\omega {\mu _e}{\bf{H}}_{sc}^{\left( \omega  \right)}}\\
{\nabla  \times {\bf{H}}_{sc}^{\left( \omega  \right)} =  + i\omega {\varepsilon _e}{\bf{E}}_{sc}^{\left( \omega  \right)}}
\end{cases} \text{in } \Omega_e \text{ , }
\end{equation}
\end{subequations}
where $\left( {{\bf{E}}_i^{\left( \omega  \right)},{\rm{ }}{\bf{H}}_i^{\left( \omega  \right)}} \right)$ denote the fields in ${\dot{\Omega} _i}$ and $\left( {{\bf{E}}_{sc}^{\left( \omega  \right)},{\rm{ }}{\bf{H}}_{sc}^{\left( \omega  \right)}} \right)$ denote the scattered fields in ${\dot{\Omega} _e}$, namely ${\bf{E}}_{sc}^{\left( \omega  \right)} = {\bf{E}}_e^{\left( \omega  \right)} - {\bf{E}}_0^{\left( \omega  \right)}$ and ${\bf{H}}_{sc}^{\left( \omega  \right)} = {\bf{H}}_e^{\left( \omega  \right)} - {\bf{H}}_0^{\left( \omega  \right)}$.  Equations (\ref{eq:MaxFF}) have to be solved with the radiation condition at infinity for the scattered fields.
Due to the symmetry of the problem, the general solution of the source-free Maxwell's equations is expressed in each homogeneous region through the SVWFs ${\bf{M}}_{mn}^{\left( J \right)}$ and ${\bf{N}}_{mn}^{\left( J \right)}$ \cite{Jackson,Varshalovich}:
\begin{subequations}
\begin{equation}
{\bf{M}}_{mn}^{\left( J \right)}\left( {kr,\theta ,\phi } \right) = z_n^{\left( J \right)}\left( {kr} \right){{\bf{X}}_{mn}}\left( {\theta ,\phi } \right),\end{equation}
\begin{equation}
{\bf{N}}_{mn}^{\left( J \right)}\left( {kr,\theta ,\phi } \right) = \frac{1}{k}\nabla  \times {\bf{M}}_{mn}^{\left( J \right)},\end{equation}
\end{subequations}
where $z_n^{\left( J \right)} = z_n^{\left( J \right)}\left( {kr} \right)$ is one of the four kinds of the spherical Bessel functions, namely Bessel function of the first kind ${j_n} = {j_n}\left( {kr} \right)$, or Bessel function of the second kind ${y_n} = {y_n}\left( {kr} \right)$, or Bessel function of the third kind (spherical Hankel functions of the first and second kind), $h_n^{\left( 1 \right)} = h_n^{\left( 1 \right)}\left( {kr} \right)$ and $h_n^{\left( 2 \right)} = h_n^{\left( 2 \right)}\left( {kr} \right)$, which we denote, respectively, with the apices $J = 1,2,3,4$. ${{\bf{X}}_{mn}} = {{\bf{X}}_{mn}}\left( {\theta ,\phi } \right)$ is a vector spherical harmonic (Appendix \ref{app:vsh}). Both the SVWFs and the vector spherical harmonics are indexed by the order $m$ and the degree $n$.

The incident plane-wave $\left( {{\bf{E}}_0^{\left( \omega  \right)},{\rm{ }}{\bf{H}}_0^{\left( \omega  \right)}} \right)$ is decomposed in Eq. (\ref{eq:IncidentField}) through the regular SVWFs, non-singular in the center of the sphere ($J = 1$), where ${E_0}$ is the amplitude of ${{\bf{E}}_0^{\left( \omega  \right)}}$, and  the coefficients $\left\{ {p_{mn}^{\left( \omega  \right)},q_{mn}^{\left( \omega  \right)}} \right\}$ are given in Appendix B for a linearly polarized state along the $x-$axis.
Also the unknown fields $\left( {{\bf{E}}_i^{\left( \omega  \right)},{\rm{ }}{\bf{H}}_i^{\left( \omega  \right)}} \right)$ in ${\Omega _i}$ ($0 \le r < R$),  are decomposed through the regular SVWFs in Eq. (\ref{eq:SVWF_FF_int}), where ${\zeta _i}\left( \omega  \right) = \sqrt {{\mu _i}/{\varepsilon _i}\left( \omega  \right)}$, ${k_i}\left( \omega  \right) = \omega \sqrt {{\varepsilon _i}\left( \omega  \right){\mu _i}} $. The unknown scattered fields $\left( {{\bf{E}}_{sc}^{\left( \omega  \right)},{\rm{ }}{\bf{H}}_{sc}^{\left( \omega  \right)}} \right)$ in ${\dot{\Omega} _e}$ ( for $R < r$) are instead decomposed in Eq. (\ref{eq:SVWF_FF_ext}) through the radiative SVWFs, satisfying the radiation condition at infinity ($J = 3$).
\begin{widetext}
\begin{subequations}
\begin{equation}
\label{eq:IncidentField}
\begin{split}
{\bf{E}}_0^{\left( \omega  \right)}\left( {r,\theta ,\phi } \right) =  - {E_0}\sum\limits_{n = 1}^\infty  {\sum\limits_{m =  - n}^n {\left\{ {q_{mn}^{\left( \omega  \right)}{\bf{M}}_{mn}^{\left( 1 \right)}\left[ {{k_e}\left( \omega  \right)r,\theta ,\phi } \right] + p_{mn}^{\left( \omega  \right)}{\bf{N}}_{mn}^{\left( 1 \right)}\left[ {{k_e}\left( \omega  \right)r,\theta ,\phi } \right]} \right\}} } \\
{\bf{H}}_0^{\left( \omega  \right)}\left( {r,\theta ,\phi } \right) = \frac{{{E_0}}}{{i{\zeta _e}}}\sum\limits_{n = 1}^\infty  {\sum\limits_{m =  - n}^n {\left\{ {p_{mn}^{\left( \omega  \right)}{\bf{M}}_{mn}^{\left( 1 \right)}\left[ {{k_e}\left( \omega  \right)r,\theta ,\phi } \right] + q_{mn}^{\left( \omega  \right)}{\bf{N}}_{mn}^{\left( 1 \right)}\left[ {{k_e}\left( \omega  \right)r,\theta ,\phi } \right]} \right\}} } ,
\end{split}
\end{equation}
%\end{widetext}
%\begin{widetext}
\begin{equation}
\label{eq:SVWF_FF_int}
\begin{split}
{\bf{E}}_i^{\left( \omega  \right)}\left( {r,\theta ,\phi } \right) =  - {E_0}\sum\limits_{n = 1}^\infty  {\sum\limits_{m =  - n}^n {\left\{ {c_{mn}^{\left( \omega  \right)}{\bf{M}}_{mn}^{\left( 1 \right)}\left[ {{k_i}\left( \omega  \right)r,\theta ,\phi } \right] + d_{mn}^{\left( \omega  \right)}{\bf{N}}_{mn}^{\left( 1 \right)}\left[ {{k_i}\left( \omega  \right)r,\theta ,\phi } \right]} \right\}} } \\
{\bf{H}}_i^{\left( \omega  \right)}\left( {r,\theta ,\phi } \right) = \frac{{{E_0}}}{{i{\zeta _i}\left( \omega  \right)}}\sum\limits_{n = 1}^\infty  {\sum\limits_{m =  - n}^n {\left\{ {d_{mn}^{\left( \omega  \right)}{\bf{M}}_{mn}^{\left( 1 \right)}\left[ {{k_i}\left( \omega  \right)r,\theta ,\phi } \right] + c_{mn}^{\left( \omega  \right)}{\bf{N}}_{mn}^{\left( 1 \right)}\left[ {{k_i}\left( \omega  \right)r,\theta ,\phi } \right]} \right\}} } ,
\end{split}
\end{equation}
%\end{widetext}
%\begin{widetext}
\begin{equation}
\label{eq:SVWF_FF_ext}
\begin{split}
{\bf{E}}_{sc}^{\left( \omega  \right)}\left( {r,\theta ,\phi } \right) = {E_0}\sum\limits_{n = 1}^\infty  {\sum\limits_{m =  - n}^n {\left\{ {b_{mn}^{\left( \omega  \right)}{\bf{M}}_{mn}^{\left( 3 \right)}\left[ {{k_e}\left( \omega  \right)r,\theta ,\phi } \right] + a_{mn}^{\left( \omega  \right)}{\bf{N}}_{mn}^{\left( 3 \right)}\left[ {{k_e}\left( \omega  \right)r,\theta ,\phi } \right]} \right\}} } \\
{\bf{H}}_{sc}^{\left( \omega  \right)}\left( {r,\theta ,\phi } \right) =  - \frac{{{E_0}}}{{i{\zeta _e}}}\sum\limits_{n = 1}^\infty  {\sum\limits_{m =  - n}^n {\left\{ {a_{mn}^{\left( \omega  \right)}{\bf{M}}_{mn}^{\left( 1 \right)}\left[ {{k_e}\left( \omega  \right)r,\theta ,\phi } \right] + b_{mn}^{\left( \omega  \right)}{\bf{N}}_{mn}^{\left( 1 \right)}\left[ {{k_e}\left( \omega  \right)r,\theta ,\phi } \right]} \right\}} } .
\end{split}
\end{equation}
\end{subequations}
\end{widetext}
The decomposition (\ref{eq:SVWF_FF_ext}) of $\left( {{\bf{E}}_{sc}^{\left( \omega  \right)},{\bf{H}}_{sc}^{\left( \omega  \right)}} \right)$ satisfies Eq. (\ref{eq:MaxFF_ext}), and the decomposition (\ref{eq:SVWF_FF_int}) of $\left( {{\bf{E}}_i^{\left( \omega  \right)},{\bf{H}}_i^{\left( \omega  \right)}} \right)$ satisfies Eq. (\ref{eq:MaxFF_int}). The unknown coefficients $\left\{ {a_{mn}^{\left( \omega  \right)},b_{mn}^{\left( \omega  \right)}} \right\}$ and $\left\{ {c_{mn}^{\left( \omega  \right)},d_{mn}^{\left( \omega  \right)}} \right\}$ are determined by requiring that the decompositions (\ref{eq:SVWF_FF_int},\ref{eq:SVWF_FF_ext}) also satisfy the boundary conditions (\ref{eq:MaxFF_bc}). The analytical expressions of $\left\{ {a_{mn}^{\left( \omega  \right)},b_{mn}^{\left( \omega  \right)}} \right\}$ and $\left\{ {c_{mn}^{\left( \omega  \right)},d_{mn}^{\left( \omega  \right)}} \right\}$ as functions of $\left\{ {{p_{mn}^{\left( \omega  \right)}},{q_{mn}^{\left( \omega  \right)}}} \right\}$ are given in Appendix \ref{app:abcdFF}.
\subsection{\label{SHField}Electromagnetic fields at second-harmonic}
The SH electromagnetic field satisfies the Maxwell's equations:
\begin{subequations}
\begin{equation}
\label{eq:MaxSH}
\begin{cases}
{\nabla  \times {\bf{E}}_i^{\left( {2\omega } \right)} =  - 2i\omega {\mu _i}{\bf{H}}_i^{\left( {2\omega } \right)}} \\
{\nabla  \times {\bf{H}}_i^{\left( {2\omega } \right)} = {\rm{ }}2i\omega {\varepsilon _i}\left( {2\omega } \right){\bf{E}}_i^{\left( {2\omega } \right)} + {\bf{J}}_b^{\left( {2\omega } \right)}}
\end{cases}\text{ in } {\hat \Omega _i} \text{,}
\end{equation}
\begin{equation}
\begin{cases}
{{\bf{\hat n}} \times \left( {{\bf{H}}_i^{\left( {2\omega } \right)} - {\bf{H}}_e^{\left( {2\omega } \right)}} \right) = - {\bf{j}}_{elet}^{\left( {2\omega } \right)}} \\
{{\bf{\hat n}} \times \left( {{\bf{E}}_i^{\left( {2\omega } \right)} - {\bf{E}}_e^{\left( {2\omega } \right)}} \right) = {\bf{j}}_{mag}^{\left( {2\omega } \right)}}
\end{cases}\text{ on } \Sigma \text{ , }
\end{equation}
\begin{equation}
\begin{cases}
{\nabla  \times {\bf{E}}_e^{\left( {2\omega } \right)} =  - 2i\omega {\mu _e}{\bf{H}}_e^{\left( {2\omega } \right)}} \\
{\nabla  \times {\bf{H}}_e^{\left( {2\omega } \right)} = {\rm{ }}2i\omega {\varepsilon _e}{\bf{E}}_e^{\left( {2\omega } \right)}}
\end{cases}\text{ in } {\hat \Omega _e} \text{ , }
\end{equation}
\end{subequations}
where 
\begin{subequations}
\begin{equation}
\label{eq:BulkEletCur}
{\bf{J}}_b^{\left( {2\omega } \right)} = 2i\omega {\bf{P}}_b^{\left( {2\omega } \right)} ,\end{equation}
\begin{equation}
\label{eq:SurfEletCur}
{\bf{j}}_{elet}^{\left( {2\omega } \right)} =  - 2i\omega {\bf{\hat n}} \times \left( {{\bf{\hat n}} \times {\bf{P}}_s^{\left( {2\omega } \right)}} \right) ,
\end{equation}
\begin{equation}
\label{eq:SurfMagnCur}
{\bf{j}}_{mag}^{\left( {2\omega } \right)}{\rm{  = }}\frac{1}{{\varepsilon '}}{\bf{\hat n}} \times {\nabla _s}\left( {{\bf{\hat n}} \cdot {\bf{P}}_s^{\left( {2\omega } \right)}} \right),
\end{equation}
\end{subequations}
the operator ${\nabla _s}$ denotes the surface gradient, $\left( {{\bf{E}}_i^{\left( {2\omega } \right)},{\rm{ }}{\bf{H}}_i^{\left( {2\omega } \right)}} \right)$  denote the SH fields in ${\dot \Omega _i}$, $\left( {{\bf{E}}_e^{\left( {2\omega } \right)},{\rm{ }}{\bf{H}}_e^{\left( {2\omega } \right)}} \right)$ denote the SH fields in ${\dot \Omega _e}$ and $\varepsilon '$ is the \textit{selvedge region} permittivity \cite{Stegeman1980}, which we assumed equal to ${\varepsilon _0}$. The sources of the SH radiation, therefore, are of three types. The volume current density field ${\bf{J}}_b^{\left( {2\omega } \right)}$ given by Eq. (\ref{eq:BulkEletCur}), takes into account the contribution of the SH bulk nonlinear polarization. The surface electric current density ${\bf{j}}_{elet}^{\left( {2\omega } \right)}$ given by Eq. (\ref{eq:SurfEletCur}), takes into account the contribution of the SH tangent surface nonlinear polarization. The surface magnetic current density ${\bf{j}}_{mag}^{\left( {2\omega } \right)}$ given by (\ref{eq:SurfMagnCur}), takes into account the contribution of the SH normal surface nonlinear polarization (see Appendix D). The systems of Eq. (\ref{eq:MaxSH}) have to be solved with the radiation condition at infinity for $\left( {{\bf{E}}_e^{\left( {2\omega } \right)},{\rm{ }}{\bf{H}}_e^{\left( {2\omega } \right)}} \right)$.

The SH field equations are formally the same of the fundamental field equations except for the bulk source term ${\bf{J}}_b^{\left( {2\omega } \right)}$ in the Maxwell-Ampere equation and the substitutions $\omega  \to 2\omega $. Consequently, the problem is reduced to that already solved for the fundamental fields, by expressing the electromagnetic field inside the nanoparticle ($\Omega_i$) as:
\begin{equation}
\begin{cases}
{\bf{E}}_i^{\left( {2\omega } \right)} =
{\bf{E}}_{hom}^{\left( {2\omega } \right)} + {\bf{E}}_{par}^{\left( {2\omega } \right)} \\
{\bf{H}}_i^{\left( {2\omega } \right)} =
{\bf{H}}_{hom}^{\left( {2\omega } \right)} + {\bf{H}}_{par}^{\left( {2\omega } \right)}
\end{cases}
\text{ , }
\end{equation}
where $\left( {{\bf{E}}_{hom}^{\left( {2\omega } \right)},{\rm{ }}{\bf{H}}_{hom}^{\left( {2\omega } \right)}} \right)$ is the general solution of Eq. (\ref{eq:MaxSH}) in absence of the source term, and $\left( {{\bf{E}}_{par}^{\left( {2\omega } \right)},{\rm{ }}{\bf{H}}_{par}^{\left( {2\omega } \right)}} \right)$ is a particular solution of the complete system of Eq. (\ref{eq:MaxSH}).
The contribution $\left( {{\bf{E}}_{hom}^{\left( {2\omega } \right)},{\rm{ }}{\bf{H}}_{hom}^{\left( {2\omega } \right)}} \right)$ will be represented as the electromagnetic field at the fundamental frequency.
The particular solution $\left( {{\bf{E}}_{par}^{\left( {2\omega } \right)},{\rm{ }}{\bf{H}}_{par}^{\left( {2\omega } \right)}} \right)$ contains two contributions, one takes into account the $\gamma$ term $\left( {{\bf{E}}_{\gamma}^{\left( {2\omega } \right)},{\rm{ }}{\bf{H}}_{\gamma}^{\left( {2\omega } \right)}} \right)$ and the other takes into account the $\delta'$ term $\left( {{\bf{E}}_{\delta'}^{\left( {2\omega } \right)},{\rm{ }}{\bf{H}}_{\delta'}^{\left( {2\omega } \right)}} \right)$.
The term $\left( {{\bf{E}}_{\gamma}^{\left( {2\omega } \right)},{\rm{ }}{\bf{H}}_{\gamma}^{\left( {2\omega } \right)}} \right)$ is given by the simple expression \cite{Heinz2004}:
\begin{equation}\label{eq:Egamma}
\begin{cases}
{\bf{E}}_{\gamma}^{\left( {2\omega } \right)} =  - \frac{{\varepsilon _0}}{{{\varepsilon _i}\left( {2\omega } \right)}} \gamma \nabla \left( {{{\bf{E}}^{\left( \omega  \right)}} \cdot {{\bf{E}}^{\left( \omega  \right)}}} \right) \\
{\bf{H}}_{\gamma}^{\left( {2\omega } \right)} =  0
\end{cases}
\end{equation}
Instead, the term $\left( {{\bf{E}}_{\delta'}^{\left( {2\omega } \right)},{\rm{ }}{\bf{H}}_{\delta'}^{\left( {2\omega } \right)}} \right)$ can be evaluated by using the Green's function for a medium with electric permittivity $\varepsilon_i$ and magnetic permeability $\mu_i$:
\begin{equation}
\begin{cases}
{\bf{E}}_{\delta'}^{(2\omega)} =
-i \omega \mu_0 &\iiint_{\Omega_i} \bm{J}_{\delta'}^{(2\omega)}(\bm{r}') g_i^{(2\omega)}(\bm{r}-\bm{r}') dV' \\
~~~~~~~~-\frac{1}{\varepsilon_i(2\omega)}\nabla
&\left[
\iiint_{\Omega_i} \rho_{\delta'}^{(2\omega)}(\bm{r}') g_i^{(2\omega)}(\bm{r}-\bm{r}') dV'
\right.
\\
&+\left.
\iint_{\Sigma_i} \eta_{\delta'}^{(2\omega)}(\bm{r}') g_i^{(2\omega)}(\bm{r}-\bm{r}') dS'
\right]
\\
{\bf{H}}_{\delta'}^{(2\omega)} =
~~~~~~
\nabla \times &\iiint_{\Omega_i} \bm{J}_{\delta'}^{(2\omega)}(\bm{r}') g_i^{(2\omega)}(\bm{r}-\bm{r}') dV'
\end{cases}
\end{equation}
where the volumetric current density $\bm{J}_{\delta'}^{(2\omega)}$, the volumetric charge density $\rho_{\delta'}^{(2\omega)}$ and the surface charge density $\eta_{\delta'}^{(2\omega)}$ are given by:
\begin{equation}
\begin{aligned}
\bm{J}_{\delta'}^{(2\omega)}=i2\omega [\varepsilon_0 \delta' (\bm{E}_i^{(\omega)}\cdot\nabla)\bm{E}_i^{(\omega)}] \\
\rho_{\delta'}^{(2\omega)}=-\nabla\cdot [\varepsilon_0 \delta' (\bm{E}_i^{(\omega)}\cdot\nabla)\bm{E}_i^{(\omega)}] \\
\eta_{\delta'}^{(2\omega)}=\bm{\hat n} \cdot [\varepsilon_0 \delta' (\bm{E}_i^{(\omega)}\cdot\nabla)\bm{E}_i^{(\omega)}]
\end{aligned}
\end{equation}
and $g_i^{(2\omega)}(\bm{r}-\bm{r}')=\frac{e^{-ik_i(2\omega)|\bm{r}-\bm{r}'|}}{4\pi|\bm{r}-\bm{r}'|}$ is the homogeneous space Green's function inside the particle.
Therefore, the fields $\left( {{\bf{E}}_{hom}^{\left( {2\omega } \right)},{\rm{ }}{\bf{H}}_{hom}^{\left( {2\omega } \right)}} \right)$ and $\left( {{\bf{E}}_e^{\left( {2\omega } \right)},{\rm{ }}{\bf{H}}_e^{\left( {2\omega } \right)}} \right)$ are  solutions of the homogeneous Maxwell's equations at frequency $2\omega $ and have to satisfy on $\Sigma$ the boundary equations:
\begin{equation}
\label{eq:Max_SH_bc}
\begin{cases}
{{\bf{\hat r}} \times \left( {{\bf{H}}_{hom}^{\left( {2\omega } \right)} - {\bf{H}}_e^{\left( {2\omega } \right)}} \right) = - {\bf{j}}_{elet}^{\left( {2\omega } \right)} - {\bf{\hat r}} \times {\bf{H}}_{\delta'}^{\left( {2\omega }  \right)}} \\
{{\bf{\hat r}}\times \left( {{\bf{E}}_{hom}^{\left( {2\omega } \right)} - {\bf{E}}_e^{\left( {2\omega } \right)}} \right) = ( {\bf{j}}_{mag}^{\left( {2\omega } \right)} - {\bf{\hat r}} \times {\bf{E}}_{\gamma}^{\left( {2\omega }  \right)} ) - {\bf{\hat r}} \times {\bf{E}}_{\delta'}^{\left( {2\omega }  \right)} }
\end{cases}
\end{equation}
By combining Equations (\ref{eq:SurfMagnCur}), (\ref{eq:Egamma}) and (\ref{eq:Max_SH_bc}), it results that the contribution of the term $\gamma$ to the SH electromagnetic field at the external of the particle may be described by the equivalent surface sources $\chi_{eff}\left[ \bm{\hat n}\bm{\hat n}\bm{\hat n} +\bm{\hat n}(\bm{\hat t}_1 \bm{\hat t}_1 + \bm{\hat t}_2 \bm{\hat t}_2 ) \right]$, where $\chi_{eff}=\gamma(\omega) \varepsilon_0 / \varepsilon_i(2\omega)$.
%$\hat{\bm{n}}\hat{\bm{t}}_1\hat{\bm{t}}_1[\gamma(\omega) \varepsilon_0 / \varepsilon_i(2\omega)]$, 
%$\hat{\bm{n}}\hat{\bm{t}}_2\hat{\bm{t}}_2[\gamma(\omega) \varepsilon_0 / \varepsilon_i(2\omega)]$ and
%$\hat{\bm{n}}\hat{\bm{n}}  \hat{\bm{n}}  [\gamma(\omega) \varepsilon_0 / \varepsilon_i(2\omega)]$
%defined on the nanoparticle surface.
%It is worth to notice that this equivalence has been derived without any reference to the shape of the particle.
%This provides a very simple demonstration of the inference made by Sipe et al. in Ref \cite{Stegeman1980}.
%According to the discussion in the Introduction, we neglect the contribution of $\delta'$ term.
%This extends to any shape the results obtained by Sipe in Refs. \onlinecite{Stegeman1980,Mizrahi1988}.

The unknown fields $\left( {{\bf{E}}_{hom}^{\left( {2\omega } \right)},{\bf{H}}_{hom}^{\left( {2\omega } \right)}} \right)$ and $\left( {{\bf{E}}_e^{\left( {2\omega } \right)},{\bf{H}}_e^{\left( {2\omega } \right)}} \right)$ are represented as:
\begin{widetext}
\begin{subequations}
\label{eq:SVWF_SH}
\begin{equation}
\begin{aligned}
{\bf{E}}_{hom}^{\left( {2\omega } \right)}\left( {r,\theta ,\phi } \right) =  - E_c^{\left( {2\omega } \right)}\sum\limits_{n = 1}^\infty  {\sum\limits_{m =  - n}^n {\left\{ {c_{mn}^{\left( {2\omega } \right)}{\bf{M}}_{mn}^{\left( 1 \right)}\left[ {{k_i}\left( {2\omega } \right)r,\theta ,\phi } \right] + d_{mn}^{\left( {2\omega } \right)}{\bf{N}}_{mn}^{\left( 1 \right)}\left[ {{k_i}\left( {2\omega } \right)r,\theta ,\phi } \right]} \right\}} } \\
{\bf{H}}_{hom}^{\left( {2\omega } \right)}\left( {r,\theta ,\phi } \right) = \frac{{E_c^{\left( {2\omega } \right)}}}{{i{\zeta _i}\left( {2\omega } \right)}}\sum\limits_{n = 1}^\infty  {\sum\limits_{m =  - n}^n {\left\{ {d_{mn}^{\left( {2\omega } \right)}{\bf{M}}_{mn}^{\left( 1 \right)}\left[ {{k_i}\left( {2\omega } \right)r,\theta ,\phi } \right] + c_{mn}^{\left( {2\omega } \right)}{\bf{N}}_{mn}^{\left( 1 \right)}\left[ {{k_i}\left( {2\omega } \right)r,\theta ,\phi } \right]} \right\}} }
\end{aligned}
\text{ for } r < R \text{, and }
\end{equation}
%for $0 \le r < R$, and
\begin{equation}
\begin{aligned}
{\bf{E}}_e^{\left( {2\omega } \right)}\left( {r,\theta ,\phi } \right) = E_c^{\left( {2\omega } \right)}\sum\limits_{n = 1}^\infty  {\sum\limits_{m =  - n}^n {\left\{ {b_{mn}^{\left( {2\omega } \right)}{\bf{M}}_{mn}^{\left( 3 \right)}\left[ {{k_e}\left( {2\omega } \right)r,\theta ,\phi } \right] + a_{mn}^{\left( {2\omega } \right)}{\bf{N}}_{mn}^{\left( 3 \right)}\left[ {{k_e}\left( {2\omega } \right)r,\theta ,\phi } \right]} \right\}} } \\
{\bf{H}}_e^{\left( {2\omega } \right)}\left( {r,\theta ,\phi } \right) =  - \frac{{E_c^{\left( {2\omega } \right)}}}{{i{\zeta _e}}}\sum\limits_{n = 1}^\infty  {\sum\limits_{m =  - n}^n {\left\{ {a_{mn}^{\left( {2\omega } \right)}{\bf{M}}_{mn}^{\left( 1 \right)}\left[ {{k_e}\left( {2\omega } \right)r,\theta ,\phi } \right] + b_{mn}^{\left( {2\omega } \right)}{\bf{N}}_{mn}^{\left( 1 \right)}\left[ {{k_e}\left( {2\omega } \right)r,\theta ,\phi } \right]} \right\}} } 
\end{aligned}
\text{ for } r > R ,
\end{equation}
\end{subequations}
\end{widetext}
%for $r > R$,
where $E_c^{\left( {2\omega } \right)} = {{{E_0}^2}}/{{{E_b}}}$
%\begin{equation}
%E_c^{\left( {2\omega } \right)} =
%{{{\left( {\frac{{{\lambda}}}{{{\lambda _p}}}} \right)}}\frac{{{E_0}}}{{{E_b}}}}
%{\chi _b}\left( \omega  \right){E_0}
%\end{equation}
is a characteristic electric field expressing the order of magnitude of the SH electric field in the nanoparticle, and
$E_b= {\omega c_0 m_{eff}}/{( {\chi _b}(\omega) e)}$.
%$E_b= {(e n_0 c_0 \omega)}/{(\varepsilon_0 {\chi _b}(\omega) \omega_p^2)}$.
%$E_b= \frac{e n_0 c_0 \omega}{\varepsilon_0 {\chi _b}\left( \omega  \right) \omega_p^2}$.
%$E_b= \frac{1}{{\chi _b}\left( \omega  \right)} \frac{e n_0 c_0}{\varepsilon_0} \frac{\omega}{\omega_p^2}$.
Both the expressions of $\left( {{\bf{E}}_{hom}^{\left( {2\omega } \right)},{\bf{H}}_{hom}^{\left( {2\omega } \right)}} \right)$ and $\left( {{\bf{E}}_e^{\left( {2\omega } \right)},{\bf{H}}_e^{\left( {2\omega } \right)}} \right)$ satisfy the homogeneous Maxwell's equations at frequency $2\omega $. The unknown coefficients $\left\{ {a_{mn}^{\left( {2\omega } \right)},b_{mn}^{\left( {2\omega } \right)}} \right\}$ and $\left\{ {c_{mn}^{\left( {2\omega } \right)},d_{mn}^{\left( {2\omega } \right)}} \right\}$ are evaluated by imposing the boundary equations (\ref{eq:Max_SH_bc}).

When the contribution of the $\delta'$ term is negligible, the right-hand sides of Eq. (\ref{eq:Max_SH_bc}) are given, in terms of SVWFs, by:
% $- {\bf{\hat r}} \times {\bf{E}}_{part}^{\left( {2\omega } \right)}\left| {_\Sigma } \right.$ in the system of Eq. (\ref{eq:Max_SH_bc}), playing the same role as the surface magnetic current ${\bf{j}}_{mag}^{\left( {2\omega } \right)}$. In order to analytically evaluate the coefficients $\left\{ {a_{mn}^{\left( {2\omega } \right)},b_{mn}^{\left( {2\omega } \right)}} \right\}$ and $\left\{ {c_{mn}^{\left( {2\omega } \right)},d_{mn}^{\left( {2\omega } \right)}} \right\}$, the fields ${\bf{j}}_{elet}^{\left( {2\omega } \right)}\left( {\theta ,\phi } \right)$ and ${\bf{j}}_{mag}^{\left( {2\omega } \right)}\left( {\theta ,\phi } \right) - {\bf{\hat r}} \times {\bf{E}}_{part}^{\left( {2\omega } \right)}\left( {r = R,\theta ,\phi } \right)$ are expanded in terms of SVWFs:
\begin{widetext}
\begin{equation}
\begin{aligned}
-&{\bf{j}}_{elet}^{\left( {2\omega } \right)}\left( {\theta ,\phi } \right) = \frac{{E_c^{\left( {2\omega } \right)}}}{{i\zeta _e}}\sum\limits_{n = 1}^\infty  {\sum\limits_{m =  - n}^n {\left[ {{v'}_{mn}^{\left( {2\omega } \right)}{{\bf{X}}_{mn}}\left( {\theta ,\phi } \right) + {u'}_{mn}^{\left( {2\omega } \right)}{\bf{\hat r}} \times {{\bf{X}}_{mn}}\left( {\theta ,\phi } \right)} \right]} } \text{ , } \\
&{\bf{j}}_{mag}^{\left( {2\omega } \right)}{\rm{ }}\left( {\theta ,\phi } \right) - {\bf{\hat r}} \times {\bf{E}}_{part}^{\left( {2\omega } \right)}\left( {r = R,\theta ,\phi } \right) =  - E_c^{\left( {2\omega } \right)}\sum\limits_{n = 1}^\infty  {\sum\limits_{m =  - n}^n \left[{{{u''}_{mn}^{\left( {2\omega } \right)}{\bf{\hat r}} \times \left[ {{\bf{\hat r}} \times {{\bf{X}}_{mn}}\left( {\theta ,\phi } \right)} \right]} } \right] } \text{ , }
\end{aligned}
\end{equation}
%\end{widetext}
where the coefficients $\left\{ {u{'}_{mn}^{\left( {2\omega } \right)},v{'}_{mn}^{\left( {2\omega } \right)}} \right\}$ and $\left\{ {u{''}_{mn}^{\left( {2\omega } \right)}},{v{''}_{mn}^{\left( {2\omega } \right)}} \right\}$  are evaluated in Appendix \ref{app:uvSH}.
The formulas of the unknown coefficients $\left\{ a_{mn}^{\left( {2\omega } \right)}, b_{mn}^{\left( {2\omega } \right)} \right\}$ and $\left\{ c_{mn}^{\left( {2\omega } \right)}, d_{mn}^{\left( {2\omega } \right)} \right\}$ are given in Appendix \ref{app:abcdSH}.
\end{widetext}

\section{\label{results}Discussion: gold nanosphere}
In the present Section, by using the analytical solution derived in the previous one, we analyze the SH generation from an isolated gold nanosphere in vacuum, as the radius, the pump wavelength and polarization vary.
Specifically, we study the SH radiation generated at pump wavelengths of ${\lambda} = 780~nm$ (Ti:sapphire laser) and ${\lambda} = 520~nm$ (gold plasmon resonance). Particular care has been devoted to the comparison with the existing theories in the Rayleigh regime. In order to model the bulk linear susceptibility of gold, we interpolated Johnson and Christy's experimental data \cite{Christy1972}.
In order to adequately represent the electromagnetic fields at the fundamental and the second-harmonic frequencies, it has been sufficient to consider the degree $n$ up to $10$, for the cases of our interest. Only the first $3$ and the first $6$ multipoles have significant amplitude at the fundamental and the second-harmonic frequencies, respectively.
Following Ref. \onlinecite{Brevet2010PRB}, we express ${\chi _{ \bot  \bot  \bot }}$, ${\chi _{\parallel  \bot \parallel }}$ and $\gamma$ in terms of the \textit{Rudnick-Stern (R-S) parameters} $\left( {a,b,d} \right)$ \cite{Stern1971,Stegeman1980}:
%Following Ref. [\onlinecite{Brevet2010PRB},\onlinecite{Stegeman1980}], we express the parameters ${\chi _{ \bot  \bot  \bot }}$, ${\chi _{\parallel  \bot \parallel }}$ and $\gamma$ through the dimensionless phenomenological \textit{Rudnick-Stern (R-S) parameters} $\left( {a,b,d} \right)$, \cite{Stern1971} as
\begin{subequations}
\label{eq:RSmodel}
\begin{align}
{\chi _{ \bot  \bot  \bot }} =  - \frac{a}{4} &{\chi_b}\left( \omega  \right)\frac{{\omega _p^2}}{{{\omega ^2}}}\frac{{\varepsilon _0}}{{e{n_0}}}, \\
{\chi _{\parallel  \bot \parallel }} =  - \frac{b}{2} &{\chi_b}\left( \omega  \right)\frac{{\omega _p^2}}{{{\omega ^2}}}\frac{{\varepsilon _0}}{{e{n_0}}}, \\
\gamma  =  - \frac{d}{8} &{\chi_b}\left( \omega  \right)\frac{{\omega _p^2}}{{{\omega ^2}}}\frac{{\varepsilon _0}}{{e{n_0}}},
\end{align}
\end{subequations}
where ${\chi _b}$ is the bulk linear susceptibility of the metal.
By choosing $\left( {a = 1,~b =  - 1,~d = 1} \right)$, we obtain the Sipe's hydrodynamic model \cite{Stegeman1980}. By measuring the SH generated by gold spherical nanoparticles with $R = 150~nm$ at $\lambda = 800~nm$, Bachelier et al. have found that an optimal choice for the phenomenological parameters $a$, $b$ and $d$ should be $\left( {a = 0.5 - i0.25,~b = 0.1,~d = 1} \right)$ \cite{Brevet2010PRB}. We discuss the solutions obtained by using both sets of values.

\begin{figure}
\includegraphics{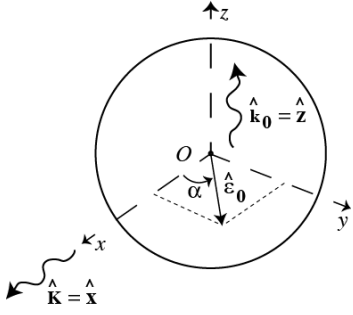}%
\caption{\label{figure02}
SH scattering geometry. The pump electromagnetic field propagates along the positive direction of the $z-$axis and is linearly polarized along ${{\bm{\hat \bm \varepsilon }}_0}$. The SH scattered field is observed along the direction ${\bf{\hat K}={\bf \hat x}}$ (the scattering plane is $xOz$). The $z-$ and $y-$ components of the SH field are considered, respectively parallel ($\parallel$) and orthogonal ($\bot$) with respect to the scattering plane.}
\end{figure}

The pump electromagnetic field is a plane-wave propagating along the positive direction of the $z-$axis, and linearly polarized in the $xy$ plane, with a polarization direction ${{\bm{\hat \varepsilon }}_0}$. We indicate with $\alpha$ the angle between the unit vector ${{\bm{\hat \varepsilon }}_0}$ and $x-$axis (the reference versus is counter-clockwise, seen from the half-space $z > 0$), as shown in Fig. \ref{figure02}.

In order to characterize the SH radiation, we consider both the SH power per unit solid angle and the SH scattering cross-section.
The SH power per unit solid angle ${{dP_{{\bf{\hat \varepsilon }}^*}^{\left( {2\omega } \right)}\left( {{\bf{\hat K}}} \right)}}/{{d\Omega }}$, radiated in the farfield along the direction ${\bf{\hat K}}$ and collected by an analyzer with polarization state ${{\bm{\hat \varepsilon }}^*}$ is defined as:
\begin{equation}
\frac{{dP_{{\bf{\hat \varepsilon }}^*}^{\left( {2\omega } \right)}\left( {{\bf{\hat K}}} \right)}}{{d\Omega }} =
\mathop {\lim }\limits_{r \to \infty } \left[ {\frac{{{r^2}}}{{2{\zeta _e}}}{{\left| {{{{\bf{\hat \varepsilon }}}^*} \cdot {\bf{E}}_e^{\left( {2\omega } \right)}\left( {{\bf{\hat K}}} \right)} \right|}^2}} \right]
\text{ .}
\end{equation}
The SH scattering cross-section ${C^{\left( {2\omega } \right)}_{sca}}$ is given by:
\begin{equation}
{C^{\left( {2\omega } \right)}_{sca}} =
\mathop {\lim }\limits_{\rho \to \infty }
 \frac{{ \int_{\Sigma_{\rho}} {{\left| {{\bf{E}}_e^{(2\omega )}} \right|}^2 \cdot {\bf{\hat n}}~d\Sigma } }}{{\left| {{\bf{E}}_0^{(\omega )}} \right|}^2},
\end{equation}
where ${\Sigma_{\rho}}$ is a spherical surface with radius $\rho$, centered at the origin of the coordinate system. ${{dP_{{\bf{\hat \varepsilon }}^*}^{\left( {2\omega } \right)}\left( {{\bf{\hat K}}} \right)}}/{{d\Omega }}$ depends on the collection direction ${\bf{\hat K}}$ of the scattered SH light. ${C^{\left( {2\omega } \right)}_{sca}}$ has the physical dimensions of an area, and it is proportional to the SH generation efficiency.

In order to analyze the SH radiation polarization state, the analyzer can be polarized either parallel ($\parallel$) or perpendicular ($\bot$) to the SH scattering plane, defined by the propagation direction $\bm {\hat k_0}$ of the pump wave, and the collection direction ${\bf{\hat K}}$. We denote with
${{dP_\parallel ^{\left( {2\omega } \right)}} / {d\Omega }}$ and
${{dP_\bot      ^{\left( {2\omega } \right)}} / {d\Omega }}$
the radiated powers per unit solid angle associated to the ${\parallel}$ and ${\bot}$ components. The analysis of the polarization state of SH radiation collected at right angle from the pump beam, \textit{i.e.} $\bm{\hat K} = \bm{ \hat x}$, is very important because it allows to discriminate the radiation generated by even and odd SH multipole sources. Indeed, only the SH ${\bf{N}}_{mn}^{\left( 3 \right)}$ multipoles with odd $n$ contribute to the ${\parallel}$ component and only the SH ${\bf{N}}_{mn}^{\left( 3 \right)}$ multipoles with even $n$ contribute to the ${\bot}$ component. These behaviors are reversed for the SH ${\bf{M}}_{mn}^{\left( 3 \right)}$ multipoles.

\subsection{SH source currents: Rayleigh and Mie regimes}
\begin{figure}
\includegraphics{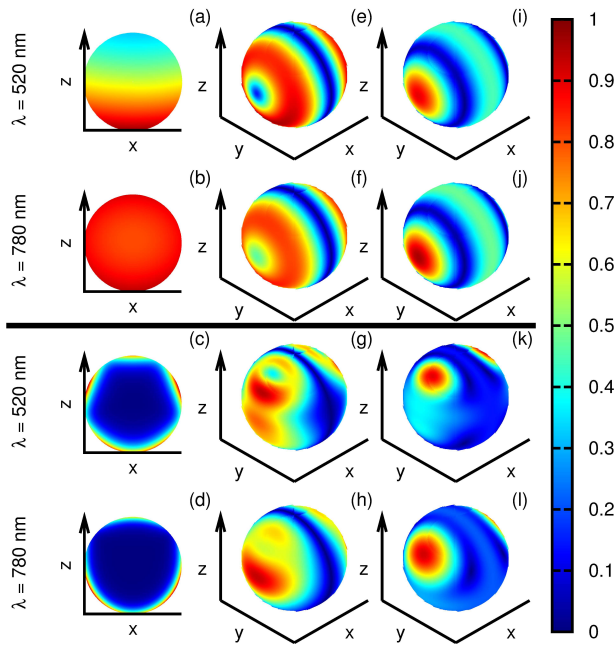}%
\caption{\label{figure03}
SH source current distribution excited by a pump field linearly polarized along $x$. The nanosphere size is $R = 10~nm$ (first and second row)and $R = 150~nm$ (third and fourth row), the pump wavelength is ${\lambda} = 520~nm$ (first and third rows) and ${\lambda} = 780~nm$ (second and fourth rows). Panels (a,b,c,d) are relative to the bulk current density cut in the $xOz$ plane, panels (e,f,g,h) are relative to the surface electric current density and panels (i,j,k,l) are relative to the surface magnetic current density. Each panel shows the current magnitude normalized to its own maximum.}
\end{figure}

Here we analyze the SH radiation generated from the single nonlinear sources, acting as if they were radiating independently. 
Fig. \ref{figure03} shows the magnitude of each SH source current density, namely ${\bf{J}}_b^{\left( {2\omega } \right)}$, ${\bf{j}}_{elet}^{\left( {2\omega } \right)}$, ${\bf{j}}_{mag}^{\left( {2\omega } \right)}$, normalized to their own maxima, and computed for the two pump wavelengths ${\lambda} = 520~nm$ and ${\lambda} = 780~nm$, corresponding to resonance and off-resonance conditions, respectively. Two nanoparticle radii have been considered, namely a particle with small size ($R = 10~nm$) and a particle comparable in size to the pump wavelength ($R = 150~nm$). The pump field is linearly polarized along the $x-$axis ($\alpha  = 0$). In particular, the first column (panels a-d) shows the magnitude of the electric current density ${\bf{J}}_b^{\left( {2\omega } \right)}$ in the $xOz$ plane, while the second and third columns (panels e-h,i-l), show the magnitude of the surface electric ${\bf{j}}_{elet}^{\left( {2\omega } \right)}$ and magnetic ${\bf{j}}_{mag}^{\left( {2\omega } \right)}$ current densities, respectively.

For small particles, ${\bf{J}}_b^{\left( {2\omega } \right)}$ is significant across the entire particle volume (panel a, b). In particular, while for ${\lambda} = 520~nm$ (panel a) ${\bf{J}}_b^{\left( {2\omega } \right)}$ decreases along the direction of forward scattering, for ${\lambda} = 780~nm$ it is almost uniform. For particles with larger size (panels c,d), the skin effect appears, \textit{i.e.} the current ${\bf{J}}_b^{\left( {2\omega } \right)}$ is strongly confined near the particle surface. The intensity distribution of both ${\bf{j}}_{elet}^{\left( {2\omega } \right)}$ and ${\bf{j}}_{mag}^{\left( {2\omega } \right)}$ (e,f,i,j) is symmetric around the polarization direction of the pump field for small particles. As the radius increases, this holds no longer true due to the onset of higher order multipoles (panels g,h,k,l). It is worth noting that the surface electric current density ${\bf{j}}_{elet}^{\left( {2\omega } \right)}$ vanishes on a circle lying in the $yOz$ plane, for any particle size and pump wavelength, as shown in panels (e-h), due to the rotational symmetry of the particle. Similarly, the surface magnetic current density ${\bf{j}}_{mag}^{\left( {2\omega } \right)}$ displays a circle with a constant magnitude for every particle size, as shown in panels (i-l).

\begin{figure}
\includegraphics{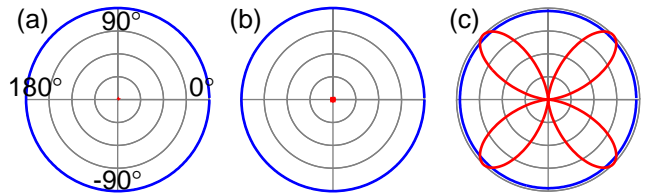}%
\caption{\label{figure04}
SH power per unit solid angle collected at right angle from the pump beam (Fig. \ref{figure02}) as function of the pump polarization angle $\alpha$, for a nanosphere with $R = 10~nm$. The pump wavelength is $780~nm$. The blue line corresponds to the $\parallel$ component and the red line to the $ \bot$ component with respect to the scattering plane. All the graphs are normalized to the maximum of the most intense component. Panel (a) is relative to the bulk current, panel (b) to the surface electric current, panel (c) to the the surface magnetic current, as if they acted separately.}
\end{figure}
In Fig. \ref{figure04} the SH power per unit solid angle collected at right angle from the pump direction is shown for a nanosphere with $R = 10~nm$. Panels (a), (b) and (c) are relevant to the SH radiation generated by the bulk, surface electric and surface magnetic SH source currents, respectively, as if they acted separately. The blue and red lines correspond respectively to
${{dP_\parallel ^{\left( {2\omega } \right)}} / {d\Omega }}$ and
${{dP_\bot      ^{\left( {2\omega } \right)}} / {d\Omega }}$
, and for each panel both the components are normalized to the maximum of the most intense.

These results agree with those obtained analyzing the SH radiation from a metal nanosphere in the Rayleigh limit \cite{Heinz1999,Heinz2004}. In this regime, the SH radiation coincides with the electromagnetic field radiated by a fictitious electric dipole with effective moment
${\bf{p}}_{eff}^{(2\omega )} ({\bf{\hat r}}) \cong {{\bf{p}}^{(2\omega )}} + i~{k_0}{{\tensor{\bf Q}^{(2\omega )}}}{\bf{\hat r}} / 3$,
where ${{\bf{p}}^{(2\omega )}}$ is the induced SH electric dipole moment ($n = 1$) and
${{\tensor{\bf Q}^{(2\omega )}}}$ is the induced SH electric quadrupole moment ($n = 2$).
Depending on the component of the SH intensity, two different shapes of the polarization diagrams arise. For each SH source,
${{dP_\parallel ^{\left( {2\omega } \right)}} / {d\Omega }}$
is due to a SH dipolar electric mode aligned along the propagation direction of the pump \cite{Heinz1999,Heinz2004}, therefore its value is independent of the polarization angle of the pump. On the other hand, the four lobe pattern observed for
${{dP_\bot ^{\left( {2\omega } \right)}} / {d\Omega }}$
 is due to a SH quadrupolar mode. Furthermore, ${{dP_\bot ^{\left( {2\omega } \right)}} / {d\Omega }}$ is negligible for both ${\bf{J}}_b^{\left( {2\omega } \right)}$ and ${\bf{j}}_{elet}^{\left( {2\omega } \right)}$, while it is comparable with ${{dP_\parallel ^{\left( {2\omega } \right)}} / {d\Omega }}$ for ${\bf{j}}_{mag}^{\left( {2\omega } \right)}$. For larger particles, higher order SH multipoles arise due to larger retardation effects, significantly modifying the SH radiation characteristics, as we shall see in the next Section.

\subsection{SH scattering cross-section}
\begin{figure}
\includegraphics{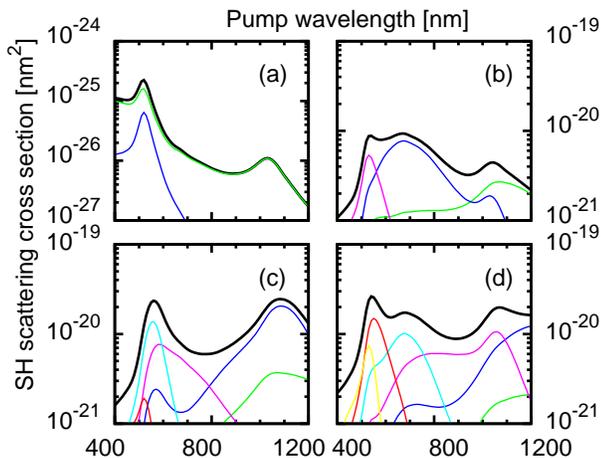}%
\caption{\label{figure05}
SH scattering cross-section (black line) as function of the pump wavelength for nanospheres with $R = 10~nm$ (a), $R = 100~nm$ (b), $R = 150~nm$ (c), $R = 200~nm$ (d), with $\left( {a = 1,b =  - 1,d = 1} \right)$. The contribution of each multipolar order to the total radiation cross-section is shown up to the 6th: $n = 1$  (green), $n = 2$ (blue), $n = 3$ (violet), $n = 4$ (cyan), $n = 5$ (red), $n = 6$ (yellow).}
\end{figure}

In this Section, we study the SH scattering cross-section $C^{\left( {2\omega } \right)}_{sca}$ for a gold nanosphere, using as \textit{R-S} parameters the values $\left( {a = 1,b =  - 1,d = 1} \right)$. All the results are relative to the case of a pump plane-wave with electric field of unitary magnitude, \textit{i.e.} $\left| E _0 \right| = 1\;V m^{-1}$ . Figure \ref{figure05} shows $C^{\left( {2\omega } \right)}_{sca}$ as function of the pump wavelength (black lines), for four particle sizes.
For all the investigated sizes, $C^{\left( {2\omega } \right)}_{sca}$ shows a maximum at $\lambda \approx 520~nm$, when the pump wavelength matches the plasmonic resonance of the particle.
Another local maximum is also observed at $\lambda \approx 1040~nm$. At this wavelength the SH fields resonate in the gold nanosphere. The relative intensity of $C^{\left( {2\omega } \right)}_{sca}$ at $\lambda \approx 1040~nm$ increases as the particle size increases.
A third local maximum can be observed at $\lambda \approx 700~nm$, for certain particle sizes (\textit{e.g.} $R = 100~nm$ and $R = 200~nm$).
 Similar trends, not shown here, have been found with the set of values $\left( {a = 0.5 - i0.25,b = 0.1,d = 1} \right)$ for the \textit{R-S} parameters.

\begin{figure}
\includegraphics{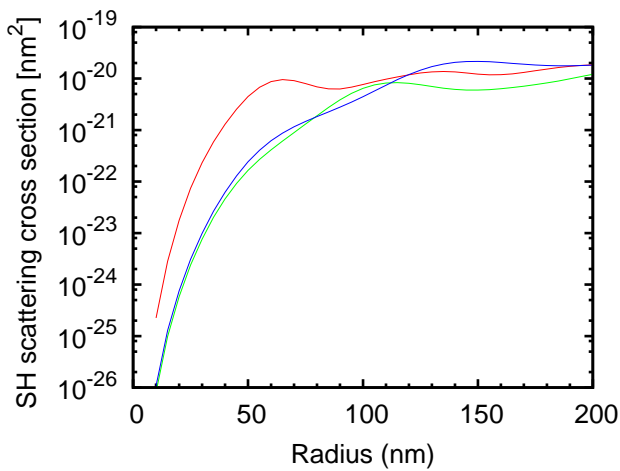}%
\caption{\label{figure06}
SH scattering cross-section as function of the nanosphere radius $R$ at pump wavelengths ${\lambda} = 520~nm$ (red), ${\lambda} = 780~nm$ (green), ${\lambda} = 1040~nm$ (blue), obtained with $\left( {a = 1,b =  - 1,d = 1} \right)$.}
\end{figure}

In order to unveil the multipolar origin of the SH radiation in the Rudnick-Stern model, the contributions of each multipole are shown, up to the $6th$ order. For $R = 10~nm$ (panel a), $C^{\left( {2\omega } \right)}_{sca}$ is mostly due to the SH dipolar source, and only for short wavelengths the SH quadrupolar source begins to be significant. As we increase the radius $R$ to $100~nm$ (panel b), we can identify three different regimes: for short wavelengths ${\lambda} < 550~nm$, $C^{\left( {2\omega } \right)}_{sca}$ is dominated by the octupolar source, the quadrupolar one prevails in the range $550~nm < {\lambda} < 950~nm$, while the dipolar source is the most intense for large wavelengths. For a particle with $R = 150~nm$ (panel c), the dipolar source is negligible regardless of the pump wavelength, and the main contributions to $C^{\left( {2\omega } \right)}_{sca}$ arise from the multipoles with $n = 2, 3, 4$. Similarly, the main contributions for a particle with $R = 200~nm$ arise from $n = 2, 3, 4$ for large wavelength, and from the multipoles $n = 5, 6$ for short wavelengths.

Figure \ref{figure06} shows the SH scattering cross-section as function of the nanoparticle radius, for three values of the pump wavelength (${\lambda} = 520~nm, 780~nm, 1040~nm$). The SH scattering cross-section increases more than 4 orders of magnitude when the particle size increases up to $R = 100~nm$. For larger radii, the SH scattering cross-section saturates and a small modulation takes place.
For small particle size, the highest cross-section is shown when the particle is in plasmonic resonance (i.e., red curve). For larger particle size, the magnitude of $C^{\left( {2\omega } \right)}_{sca}$ is comparable for all the investigated pump wavelengths. Also in this case, similar trends have been found using the set of \textit{R-S} parameters $\left( {a = 0.5 - i0.25,b = 0.1,d = 1} \right)$.

\subsection{SH power dependence on the pump polarization}
\begin{figure}
\includegraphics{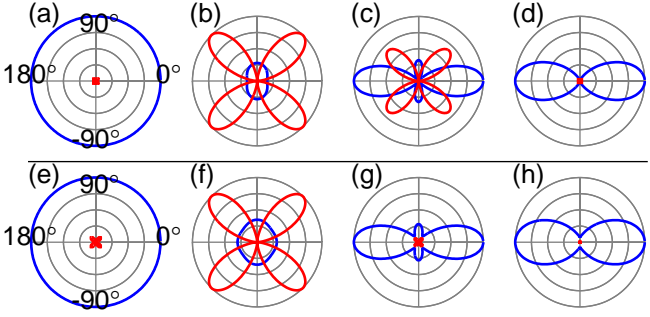}%
\caption{\label{figure07}
SH power per unit solid angle collected at right angle from the pump beam (Fig. \ref{figure02}) as function of the pump polarization angle $\alpha$, for nanosphere with $R = 10~nm$ (a,e), $R = 100~nm$ (b,f), $R = 150~nm$ (c,g), $R = 200~nm$ (d,h), obtained by using (a-d) $\left( {a = 1,b =  - 1,d = 1} \right)$ and (e-h) $\left( {a = 0.5 - i0.25,b = 0.1,d = 1} \right)$. The pump wavelength is ${\lambda} = 780~nm$. The blue line corresponds to the $\parallel$ component and the red line to the $\bot$ component with respect to the scattering plane. All the graphs are normalized to the maximum of the most intense component.}
\end{figure}

\begin{figure}
\includegraphics{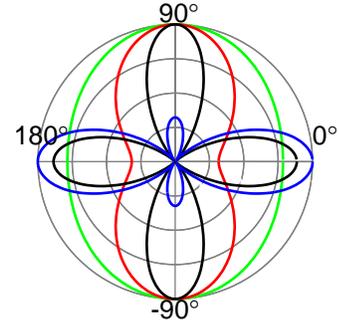}%
\caption{\label{figure08}
SH power per unit solid angle of the $\parallel$ component collected at right angle from the pump beam (Fig. \ref{figure02}), as function of the pump polarization angle $\alpha $, for nanospheres of size $R = 80~nm$ (green), $R = 120~nm$ (red), $R = 140~nm$ (black), $R = 150~nm$(blue) and $\left( {a = 1,b =  - 1,d = 1} \right)$. The pump wavelength is ${\lambda} = 780~nm$, all the curves are normalized to their own maximum.}
\end{figure}

The SH power radiated at right angle from the propagation direction of the pump allows for the recognition of even- and odd-order multipolar contributions to the SH generation process, as already pointed out.
Figure \ref{figure07} shows
${{dP_\parallel ^{\left( {2\omega } \right)}} / {d\Omega }}$ and
${{dP_\bot      ^{\left( {2\omega } \right)}} / {d\Omega }}$
 as function of the polarization angle $\alpha $ of the pump, for two different choices of \textit{R-S} parameters. The first row (a-d) is relative to the \textit{R-S} parameter set $\left( {a = 1,b =  - 1,d = 1} \right)$, while the second row (e-h) is relative to the set $\left( {a = 0.5 - i0.25,b = 0.1,d = 1} \right)$. Four different values of particle size are presented: $R = 10~nm$ (a,e), $R = 100~nm$ (b,f), $R = 150~nm$ (c,g), $R = 200~nm$ (d,h). Both
${{dP_\parallel ^{\left( {2\omega } \right)}} / {d\Omega }}$ and
${{dP_\bot      ^{\left( {2\omega } \right)}} / {d\Omega }}$  are normalized to the maximum of the most intense component.
For small (a,e) and very large (d,h) radii,
${{dP_\parallel ^{\left( {2\omega } \right)}} / {d\Omega }}$
prevails over the
${{dP_\bot      ^{\left( {2\omega } \right)}} / {d\Omega }}$
for both sets of the \textit{R-S} parameters. For intermediate sizes (b,c,f,g),
${{dP_\parallel ^{\left( {2\omega } \right)}} / {d\Omega }}$
and
${{dP_\bot      ^{\left( {2\omega } \right)}} / {d\Omega }}$
are comparable, and their relative intensities strongly depend on the particular choice of the \textit{R-S} parameters. For small particles,
${{dP_\parallel ^{\left( {2\omega } \right)}} / {d\Omega }}$
is independent of the polarization angle $\alpha $ (panels a,e), while up to 4 lobes can appear for larger particles, as it will be shown more in detail in Fig. \ref{figure08}. The graphs of the component
${{dP_\bot      ^{\left( {2\omega } \right)}} / {d\Omega }}$
feature four lobes oriented along the bisectors of the 4 quadrants, in each of the investigated case.
% Therefore the maxima for
%${{dP_\bot      ^{\left( {2\omega } \right)}} / {d\Omega }}$
%are obtained if the polarization direction of the pump and SH collection direction are tilted of $45^\circ$ or $135^\circ$.

%The appearance of an octupolar SH source generated by the retardation effects, whose intensity increases in amplitude as the particle size increases, significantly modify the characteristic of
%. In fact the interference in the far field between the SH dipolar and octupolar sources gives rise to a variety of shapes.
The appearance of an octupolar SH source significantly modifies the shape of ${{dP_\parallel ^{\left( {2\omega } \right)}} / {d\Omega }}$. In the Rayleigh limit the ${\parallel}$ component generated by the SH dipolar source fully prevails over the ${\bot}$ component generated by the SH quadrupolar source. As the radius increases, the intensity of the SH octupolar source increases due to the retardation effects, as pointed out in the previous Section. Due to the interference in the far field of the SH dipolar and octupolar fields, the ${{dP_\parallel ^{\left( {2\omega } \right)}} / {d\Omega }}$ reduces significantly until it becomes smaller than ${{dP_\bot      ^{\left( {2\omega } \right)}} / {d\Omega }}$ (b,c). As the radius further increases, the SH octupolar source prevails over the SH dipolar source, and the shape of ${{dP_\parallel ^{\left( {2\omega } \right)}} / {d\Omega }}$  gets close to that of an octupole (d,e).

The details of the transition from the dipole to the octupole pattern as the radius increases are shown in Fig. \ref{figure08}. First, the circular shape of
${{dP_\parallel ^{\left( {2\omega } \right)}} / {d\Omega }}$
is shrunk along the directions $\alpha = 0^\circ ,~180^\circ$, until the amplitude  reaches a null (green and red curves).
Then, two lobes arise along these directions (black curve), forming a four-lobe pattern. As the radius further increases, the intensities of the two lobes along the directions at $\alpha = 0^\circ,~180^\circ$ prevail over the intensities of the lobes along the directions at $\alpha  = 90^\circ,~270^\circ$ (blue curve).
In conclusion we found that, if either the SH dipole or the SH octupole prevail, the polarization properties of the SH radiation obtained by the two sets of \textit{R-S} parameters are similar. Otherwise, the interference between the two SH multipoles introduces significant differences. 
This may provide a fingerprint to evaluate the parameters in the framework of the \textit{R-S} model.

\subsection{SH radiation diagrams}
\begin{figure}
\includegraphics{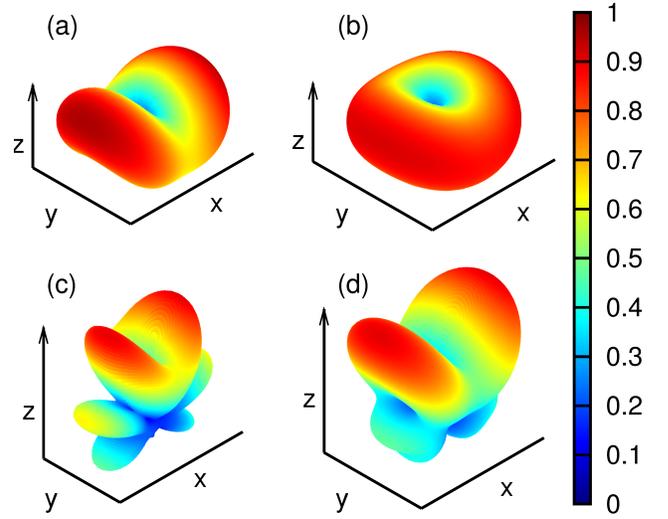}%
\caption{\label{figure09}
SH radiation diagrams for nanospheres of size $R = 10~nm$ (panels a,b), $R = 150~nm$ (panel c, d), obtained by using $\left( {a = 1,b =  - 1,d = 1} \right)$. Panels (a, c) are relative to the pump wavelength ${\lambda} = 520~nm$, and panels (b,d) to ${\lambda} = 780~nm$. All the intensities are normalized to their own maximum.}
\end{figure}

Figure \ref{figure09} shows the angular distribution of the SH radiation generated by gold nanospheres, obtained using the \textit{R-S} parameters $\left( {a = 1,~b = - 1,~d = 1} \right)$.
The first row (a,b) is relative to a small sphere ($R = 10~nm$) and the second row (c,d) to a large sphere ($R = 150~nm$). In the first column (a,c) the pump wavelength is ${\lambda} = 520~nm$, while in the second column (b,d) ${\lambda} = 780~nm$.

For particles with $R = 10~nm$, the dipolar and quadrupolar SH sources dominate the response, in agreement with the Rayleigh limit. In particular, we notice that the quadrupolar SH source is more important at ${\lambda} = 520~nm$, while at ${\lambda} = 780~nm$ the dipolar mode fully dominates the response. As the particle radius increases, higher order modes come into play, resulting in an higher number of secondary lobes. For particles with $R = 150~nm$, the octupolar mode dominates the angular distribution of the SH radiation at ${\lambda} = 780~nm$.
Moreover, for each particle size and pump wavelength, the lobes display a preferential alignment along the polarization direction of the pump field.

\begin{figure}
\includegraphics{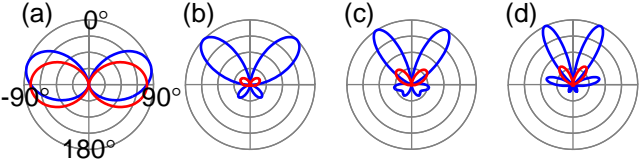}%
\caption{\label{figure10}
SH radiation diagrams as function of $\theta$ angle, at $\varphi  = 0 ^ \circ $ (blue) and $\varphi  = 90 ^ \circ $ (red), for nanospheres of size $R = 10~nm$ (a), $R = 100~nm$ (b), $R = 150~nm$ (c), and $R = 200~nm$ (d) and $\left( a = 1,b =  - 1,d = 1 \right)$. The pump wavelength is $\lambda = 780~nm$, for each panel the curves are normalized to the maximum of the most intense one.}
\end{figure}

Figure \ref{figure10} shows the two cuts of the 3D radiation diagram along the $xOz$, and $yOz$ plane.
It is worth noting that the SH power vanishes in both the forward- and backward-scattering directions, regardless of $R$. This selection rule is a direct consequence of the rotational symmetry of the sphere around the propagation direction of the pump.
Two lobes, directed orthogonally to the pump propagation direction, characterize the radiation diagrams of small particles. As the particle size increases, the lobes with higher power tend to come closer to the forward direction, while the lobes in the backward direction decrease in amplitude. The same trend is observed if the particle size is fixed and the pump wavelength decreases. This behavior has been already observed experimentally for silver nanoparticles in Ref. \onlinecite{Gonella2012}.

\section{\label{concl}Conclusions}
We have developed a full-wave analytical solution for the second-harmonic generation from metal spherical particles of arbitrary size. This approach extends the existing theories, enabling a rigorous treatment of all the sources of SH radiation, located both on the surface and in the bulk of the particle.
%We neglected the contribution of the bulk $\delta'$ source, because the main contribution to the fields at the fundamental frequency arises from low-order multipoles, in the cases of our interest. The $\delta'$ contribution may be significant when ${{\bf{E}}^{\left( \omega  \right)}}$ is rapidly varying in the bulk of the metal.
%Since the contribution of the $\delta'$ term depends on the spatial derivatives of ${{\bf{E}}^{\left( \omega  \right)}}$, \cite{Kauranen2009}
 The solution of the problem is derived in the framework of the Mie theory by expanding the pump field, the nonlinear polarization sources and the second-harmonic fields in series of spherical vector wave functions, and enforcing the boundary conditions at the sphere surface.

We investigated the SH radiation of gold nanospheres by using the Rudnick-Stern model for the SH sources.
In particular, we studied the spatial distributions of the nonlinear polarization sources, which display significant symmetries due to the spherical geometry.
We investigated the SH cross-section dependence on the pump wavelength, demonstrating the contribution of SH multipoles up to the order $N=6$, as the particle radius increases up to $\sim 200~nm$.
Similarly, we studied the multipolar origin of the SH radiation diagrams, and showed significant analogies with experimental works in literature.
Eventually, we investigated the SH radiated power as function of the pump polarization angle.
In particular, we compared the solutions obtained by using as Rudnick-Stern parameters both the Sipe's model values and the experimental values given in Ref. \onlinecite{Brevet2010PRB}.
The behavior of the SH p- and s- components strongly varies with the Rudnick-Stern parameters.
%, and the shape and the relative intensities of the SH p- and s- components are function of them.

The application of the proposed method in combination with experimental observation, can improve the general understanding of nonlinear processes in metals, and can lead to an accurate evaluation of weights for the different SH sources. The theory of SH scattering can be easily extended to the multiparticle case. This approach can also guide the design of novel nanoplasmonic devices with enhanced SH emission for a wide range of applications \cite{Lipson2004}, including sensors for probing physical and chemical properties of material surfaces.

% Specify following sections are appendices. Use \appendix* if there
% only one appendix.
\appendix

%\begin{widetext}
\section{\label{app:vsh} Vector spherical harmonics}
The vector spherical harmonics ${{\bf{X}}_{mn}}$ are \cite{Jackson,Varshalovich}:
\begin{align*}
{{\bf{X}}_{mn}}\left( {\theta ,\phi } \right) = i\frac{1}{{\sqrt {n\left( {n + 1} \right)} }}\sqrt {\frac{{2n + 1}}{{4\pi }}\frac{{\left( {n - m} \right)!}}{{\left( {n + m} \right)!}}} \cdot \\
\left[ {i{\pi _{mn}}\left( {\cos \theta } \right)\hat \theta  - {\tau _{mn}}\left( {\cos \theta } \right)\hat \phi } \right]{e^{im\phi }}
\text{ , }
\end{align*}
%\[{{\bf{X}}_{mn}}\left( {\theta ,\phi } \right) = i\frac{1}{{\sqrt {n\left( {n + 1} \right)} }}\sqrt {\frac{{2n + 1}}{{4\pi }}\frac{{\left( {n - m} \right)!}}{{\left( {n + m} \right)!}}} \left[ {i{\pi _{nm}}\left( {\cos \theta } \right)\hat \theta  - {\tau _{nm}}\left( {\cos \theta } \right)\hat \phi } \right]{e^{im\phi }}\]	(35)
%\end{widetext}
\begin{equation*}
\text{where \;\;\;}
\begin{aligned}
&{\pi _{mn}}\left( {\cos \theta } \right) &=& \frac{m}{{\sin \theta }}P_n^m\left( {\cos \theta } \right)\\
&{\tau _{mn}}\left( {\cos \theta } \right) &=& \frac{d}{{d\theta }}P_n^m\left( {\cos \theta } \right)
\end{aligned}
\text{ , }
\end{equation*}
and $P_n^m = P_n^m\left( u \right)$ is the associated Legendre function of the first kind and of degree $n$ and $m$.

\section{\label{app:pqFF}Calculation of $\left\{ {p_{mn}^{\left( \omega  \right)},q_{mn}^{\left( \omega  \right)}} \right\}$}
The expansion coefficients in Eq. (\ref{eq:IncidentField}), for a linearly polarized plane-wave propagating along the $z-$axis with the electric field parallel to the $x-$axis (Fig. \ref{figure01}a), are:
%\begin{subequations}
\begin{align*}
&p_{ mn}^{\left( \omega  \right)} = q_{mn}^{\left( \omega  \right)} = 0 \text{, for} \left| m \right| \ne 1\\
&p_{ 1n}^{\left( \omega  \right)} =
 q_{ 1n}^{\left( \omega  \right)} =
-p_{-1n}^{\left( \omega  \right)} =
 q_{-1n}^{\left( \omega  \right)} = \frac{1}{2}{{(-i)}^{n}}\sqrt{{4\pi(2n + 1)}}
\end{align*}
%\end{subequations}

\section{\label{app:abcdFF}Calculation of $\left\{ {a_{mn}^{\left( \omega  \right)},b_{mn}^{\left( \omega  \right)}} \right\}$, $\left\{ {c_{mn}^{\left( \omega  \right)},d_{mn}^{\left( \omega  \right)}} \right\}$}
The coefficients $\left\{ {a_{mn}^{\left( \omega  \right)},b_{mn}^{\left( \omega  \right)}} \right\}$ and $\left\{ {c_{mn}^{\left( \omega  \right)},d_{mn}^{\left( \omega  \right)}} \right\}$ are expressed as
\begin{equation*}
\begin{split}
\frac{{a_{mn}^{(\omega )}}}{{p_{mn}^{(\omega )}}} = \frac{{\frac{{{\zeta _e}}}{{{\zeta _i}(\omega )}}\;{\psi _n}(x_i^{(\omega )})\;{{\dot \psi }_n}(x_e^{(\omega )}) - \;{\psi _n}(x_e^{(\omega )})\;{{\dot \psi }_n}(x_i^{(\omega )})}}{{\frac{{{\zeta _e}}}{{{\zeta _i}(\omega )}}\;{\psi _n}(x_i^{(\omega )})\;{{\dot \xi }_n}(x_e^{(\omega )}) - \;{\xi _n}(x_e^{(\omega )})\;{{\dot \psi }_n}(x_i^{(\omega )})}}\\
\frac{{b_{mn}^{(\omega )}}}{{q_{mn}^{(\omega )}}} = \frac{{{\psi _n}(x_i^{(\omega )})\;{{\dot \psi }_n}(x_e^{(\omega )}) - \frac{{{\zeta _e}}}{{{\zeta _i}(\omega )}}\;{\psi _n}(x_e^{(\omega )})\;{{\dot \psi }_n}(x_i^{(\omega )})}}{{{\psi _n}(x_i^{(\omega )})\;{{\dot \xi }_n}(x_e^{(\omega )}) - \frac{{{\zeta _e}}}{{{\zeta _i}(\omega )}}\;{\xi _n}(x_e^{(\omega )})\;{{\dot \psi }_n}(x_i^{(\omega )})}}\\
\frac{{c_{mn}^{(\omega )}}}{{q_{mn}^{(\omega )}}} = \frac{{i\frac{{{k_i}(\omega )}}{{{k_e}(\omega )}}}}{{\;{\psi _n}(x_i^{(\omega )})\;{{\dot \xi }_n}(x_e^{(\omega )}) - \frac{{{\zeta _e}}}{{{\zeta _i}(\omega )}}\;{\xi _n}(x_e^{(\omega )})\;{{\dot \psi }_n}(x_i^{(\omega )})}}\\
\frac{{d_{mn}^{(\omega )}}}{{p_{mn}^{(\omega )}}} = \frac{{i\frac{{{k_i}(\omega )}}{{{k_e}(\omega )}}}}{{\frac{{{\zeta _e}}}{{{\zeta _i}(\omega )}}\;{\psi _n}(x_i^{(\omega )})\;{{\dot \xi }_n}(x_e^{(\omega )}) - \;{\xi _n}(x_e^{(\omega )})\;{{\dot \psi }_n}(x_i^{(\omega )})}}
\end{split}
\end{equation*}
where $x_e^{(\omega )} = k_e^{(\omega )}R,\;\;x_i^{(\omega )} = k_i^{(\omega )}$ and ${\psi _n} = {\psi _n}\left( \rho  \right)\;$, ${\xi _n} = {\xi _n}\left( \rho  \right)$ are the Riccati-Bessel functions defined as ${\psi _n}(\rho ) = \rho \;{j_n}(\rho ),\;\;\;{\xi _n} = \rho \;h_n^{(1)}(\rho )$. $\dot \zeta $ denotes the first derivative of $\zeta  = \zeta (\rho )$ with respect to $\rho $.

\section{\label{app:selvedge}Selvedge region}
The selvedge region (Fig. \ref{figure01}b) is a layer of infinitesimal depth $\delta $ at the interface metal-vacuum. In this region there is a volumetric current density
${\bf{J}}_{s \bot }^{\left( {2\omega } \right)} = i2\omega \left( {{\bf{P}}_s^{\left( {2\omega } \right)} \cdot {\bf{n}}/\delta } \right){\bf{n}}$,
which is exactly compensated by the normal component of the displacement current density,
${\bf{J}}_{s \bot }^{\left( {2\omega } \right)} + i2\omega {\bf{D}}_ \bot ^{\left( {2\omega } \right)} = {\bf{0}}$,
otherwise there would be an unbounded magnetic field. Therefore, in the selvedge region,
${\bf{D}}_ \bot ^{\left( {2\omega } \right)} =  - \left( {{\bf{P}}_s^{\left( {2\omega } \right)} \cdot {\bf{n}}/\delta } \right){\bf{n}}$.
From the Faraday-Neumann's law, applied to the elementary curve $\Delta l$ shown in Figure 1b, we have
$\left( {{\bf{E}}_i^{\left( {2\omega } \right)} - {\bf{E}}_e^{\left( {2\omega } \right)}} \right)\left| _{\Sigma} \right. \cdot \Delta {l_\parallel } = {u_2} - {u_1}$,
where
${u_{(\alpha)}} = \int_{\Delta l_{\bot}^{(\alpha)}}{{\bf{E}}^{\left( {2\omega } \right)}}\cdot{d{\bm l}} = \left. ({{\bf{P}}_s^{\left( {2\omega } \right)} \cdot {\bf{\hat n}}})\right| _{Q^{(\alpha)}}$
and $\alpha  = 1,2$.
By combining these relations we obtain the equations
${\bf{\hat n}} \times \left. \left( {{\bf{E}}_i^{\left( {2\omega } \right)} - {\bf{E}}_e^{\left( {2\omega } \right)}} \right)\right| {_\Sigma} = {\bf{\hat n}} \times {\nabla _S}\left( {{\bf{P}}_s^{\left( {2\omega } \right)} \cdot {\bf{\hat n}}} \right)/\varepsilon '$.

\section{\label{app:abcdSH}Calculation of $\left\{ {a_{mn}^{\left( {2\omega } \right)},b_{mn}^{\left( {2\omega } \right)}} \right\}$, $\left\{ {c_{mn}^{\left( {2\omega } \right)},d_{mn}^{\left( {2\omega } \right)}} \right\}$}
The coefficients $\left\{ {a_{mn}^{\left( {2\omega } \right)},b_{mn}^{\left( {2\omega } \right)}} \right\}$ and $\left\{ {c_{mn}^{\left( {2\omega } \right)},d_{mn}^{\left( {2\omega } \right)}} \right\}$ are expressed as:
\begin{equation*}
\begin{split}
a_{mn}^{\left( {2\omega } \right)} = {a'}_{mn}^{\left( {2\omega } \right)} + {a''}_{mn}^{\left( {2\omega } \right)} \text{ ,\;\;\; }
b_{mn}^{\left( {2\omega } \right)} = {b'}_{mn}^{\left( {2\omega } \right)} + b{''}_{mn}^{\left( {2\omega } \right)} \text{ , } \\
c_{mn}^{\left( {2\omega } \right)} = {c'}_{mn}^{\left( {2\omega } \right)} + {c''}_{mn}^{\left( {2\omega } \right)} \text{ ,\;\;\; }
d_{mn}^{\left( {2\omega } \right)} = {d'}_{mn}^{\left( {2\omega } \right)} + {d''}_{mn}^{\left( {2\omega } \right)} \text{ , } 
\end{split}
\end{equation*}
where with one apex we denote the contribution due to the tangential surface SH sources and with two apices we denote the contributions of both the normal surface SH sources and the $\gamma$ bulk SH sources.
For the contribution of the tangential surface SH sources we have:
\begin{equation*}
\begin{split}
\frac{{{a'}_{mn}^{(2\omega )}}}{{{u'}_{mn}^{(2\omega )}}} = \frac{{x_e^{(2\omega )} {{\dot \psi }_n}(x_i^{(2\omega )})}}{{{\xi _n}(x_e^{(2\omega )}) {{\dot \psi }_n}(x_i^{(2\omega )}) - \frac{{{\zeta _e}}}{{{\zeta _i}(2\omega )}}{\psi _n}(x_i^{(2\omega )}) {{\dot \xi }_n}(x_e^{(2\omega )})}}\\
%\end{split}
%\end{equation*}
%\begin{equation*}
%\begin{split}
\frac{{{b'}_{mn}^{(2\omega )}}}{{{v'}_{mn}^{(2\omega )}}} = \frac{{x_e^{(2\omega )} {\psi _n}(x_i^{(2\omega )})}}{{\frac{{{\zeta _e}}}{{{\zeta _i}(2\omega )}}{\xi _n}(x_e^{(2\omega )}) {{\dot \psi }_n}(x_i^{(2\omega )}) - {\psi _n}(x_i^{(2\omega )}) {{\dot \xi }_n}(x_e^{(2\omega )})}}\\
\frac{{{c'}_{mn}^{(2\omega )}}}{{{v'}_{mn}^{(2\omega )}}} = \frac{{x_i^{(2\omega )} {\xi _n}(x_e^{(2\omega )})}}{{{\psi _n}(x_i^{(2\omega )}) {{\dot \xi }_n}(x_e^{(2\omega )}) - \frac{{{\zeta _e}}}{{{\zeta _i}(2\omega )}}{\xi _n}(x_e^{(2\omega )}) {{\dot \psi }_n}(x_i^{(2\omega )})}}\\
\frac{{{d'}_{mn}^{(2\omega )}}}{{{u'}_{mn}^{(2\omega )}}} = \frac{{x_i^{(2\omega )} {{\dot \xi }_n}(x_e^{(2\omega )})}}{{\frac{{{\zeta _e}}}{{{\zeta _i}(2\omega )}}{\psi _n}(x_i^{(2\omega )}) {{\dot \xi }_n}(x_e^{(2\omega )}) - {\xi _n}(x_e^{(2\omega )}) {{\dot \psi }_n}(x_i^{(2\omega )})}}
\end{split}
\end{equation*}
\\
\\
where the coefficients $\left\{ {{u'}_{mn}^{(2\omega )},{v'}_{mn}^{(2\omega )}} \right\}$ are given in Appendix \ref{app:uvSH}, $x_e^{(2\omega )} = k_e^{(2\omega )}R,\;\;x_i^{(2\omega )} = k_i^{(2\omega )}R$, ${\psi _n} = {\psi _n}\left( \rho  \right)\;$, ${\xi _n} = {\xi _n}\left( \rho  \right)$ are the Riccati-Bessel functions.
% defined as ${\psi _n}(\rho ) = \rho \;{j_n}(\rho ),\;\;\;{\xi _n} = \rho \;h_n^{(1)}(\rho )$ and $\dot \zeta$ denotes the first derivative of $\zeta  = \zeta \left( \rho  \right)$ with respect to $\rho$.
For the contribution of both the normal surface SH sources and the $\gamma$ bulk SH sources we have:
\begin{equation*}
\begin{split}
&\frac{{{a''}_{mn}^{(2\omega )}}}{{{u''}_{mn}^{(2\omega )}}} = \frac{{\frac{{{\zeta _e}}}{{{\zeta _i}(2\omega )}}x_e^{(2\omega )} {\psi _n}(x_i^{(2\omega )})}}{{\frac{{{\zeta _e}}}{{{\zeta _i}(2\omega )}}{\psi _n}(x_i^{(2\omega )}) {{\dot \xi }_n}(x_e^{(2\omega )}) - {\xi _n}(x_e^{(2\omega )}) {{\dot \psi }_n}(x_i^{(2\omega )})}} \\
&{b''}_{mn}^{(2\omega )} = 0 \\
&{c''}_{mn}^{(2\omega )} = 0 \\
&\frac{{{d''}_{mn}^{(2\omega )}}}{{{u''}_{mn}^{(2\omega )}}} = \frac{{x_i^{(2\omega )} {\xi _n}(x_e^{(2\omega )})}}{{{\xi _n}(x_e^{(2\omega )}) {{\dot \psi }_n}(x_i^{(2\omega )}) - \frac{{{\zeta _e}}}{{{\zeta _i}(2\omega )}}{\psi _n}(x_i^{(2\omega )}) {{\dot \xi }_n}(x_e^{(2\omega )})}}
\end{split}
\end{equation*}
where the coefficients $\left\{ {{u''}_{mn}^{(2\omega )},{v''}_{mn}^{(2\omega )}} \right\}$ are given in Appendix \ref{app:uvSH}.

\begin{widetext}
%\onecolumngrid
\section{\label{app:uvSH}
Calculation of $\left\{ {{u'}_{mn}^{(2\omega )},{v'}_{mn}^{(2\omega )}} \right\},\left\{ {{u''}_{mn}^{(2\omega )},{v''}_{mn}^{(2\omega )}} \right\}$
}
The coefficients $\left\{ {{u'}_{mn}^{(2\omega )},{v'}_{mn}^{(2\omega )}} \right\}$ for the surface tangential source can be expressed as:
\begin{equation*}
\begin{split}
{u'}_{mn}^{(2\omega )} = i 2 \left( - \frac{b}{2} \right) \frac{{{\zeta _e}}}{{{\zeta _0}}} \sum\limits_{{n_1}}^\infty  {\sum\limits_{{m_1} =  - {n_1}}^{{n_1}} {\sum\limits_{{n_2}}^\infty  {\sum\limits_{{m_2} =  - {n_2}}^{{n_2}} {A_{{m_1}{n_1}}^{(1)}A_{{m_2}{n_2}}^{( - 1)}C_{{n_1}{m_1}{n_2}{m_2}nm}^{(1,0, - 1)} + A_{{m_1}{n_1}}^{(0)}A_{{m_2}{n_2}}^{( - 1)}C_{{n_1}{m_1}{n_2}{m_2}nm}^{(1,1, - 1)}} } } }\\
{v'}_{mn}^{(2\omega )} = - 2 \left( - \frac{b}{2} \right) \frac{{{\zeta _e}}}{{{\zeta _0}}} \sum\limits_{{n_1}}^\infty  {\sum\limits_{{m_1} =  - {n_1}}^{{n_1}} {\sum\limits_{{n_2}}^\infty  {\sum\limits_{{m_2} =  - {n_2}}^{{n_2}} {A_{{m_1}{n_1}}^{(1)}A_{{m_2}{n_2}}^{( - 1)}C_{{n_1}{m_1}{n_2}{m_2}nm}^{(0,0, - 1)} + A_{{m_1}{n_1}}^{(0)}A_{{m_2}{n_2}}^{( - 1)}C_{{n_1}{m_1}{n_2}{m_2}nm}^{(0,1, - 1)}} } } }
\end{split}
\end{equation*}
\begin{equation*}
\text{where}
\begin{split}
&A_{mn}^{(0)} = {\left. {{j_n}({k_i}(\omega )r)} \right|_{r = R}}c_{mn}^{(\omega )}\\
&A_{mn}^{(1)} = {\left. {i\frac{1}{{{k_i}(\omega )}}\left( {\frac{\partial }{{\partial r}} + \frac{1}{r}} \right)~~{j_n}\left( {{k_i}(\omega )r} \right)} \right|_{r = R}}d_{mn}^{(\omega )}\\
&A_{mn}^{( - 1)} = {\left. {i\sqrt {n(n + 1)} \frac{1}{{{k_i}(\omega )r}}~~{j_n}\left( {{k_i}(\omega )r} \right)} \right|_{r = R}}d_{mn}^{(\omega )}\\
\end{split}
\end{equation*}

\begin{equation*}\begin{split}
&C^{(0,1,-1)}_{J_1 M_1 J_2 M_2 J M} =
\sqrt{\frac{3}{2 \pi}} C^{J M}_{J_1 M_1 J_2 M_2} \cdot \\
&\left[ 
\sqrt{(J_1+1)(J_2)(2 J_1-1)(2 J_2-1)}
\left\{ \begin{matrix}J_1 & J_1-1 & 1 \\ J_2 & J_2-1 & 1 \\ J & J & 1\end{matrix} \right\}
C^{J 0}_{(J_1-1)(0)(J_2-1)(0)} \right. \\
&\left.
-\sqrt{(J_1+1)(J_2+1)(2 J_1-1)(2 J_2+3)}
\left\{ \begin{matrix}J_1 & J_1-1 & 1 \\ J_2 & J_2+1 & 1 \\ J & J & 1\end{matrix} \right\}
C^{J 0}_{(J_1-1)(0)(J_2+1)(0)} \right. \\
&\left.
+\sqrt{(J_1)(J_2)(2 J_1+3)(2 J_2-1)}
\left\{ \begin{matrix}J_1 & J_1+1 & 1 \\ J_2 & J_2-1 & 1 \\ J & J & 1\end{matrix} \right\}
C^{J 0}_{(J_1+1)(0)(J_2-1)(0)} \right. \\
&\left.
-\sqrt{(J_1)(J_2+1)(2 J_1+3)(2 J_2+3)}
\left\{ \begin{matrix}J_1 & J_1+1 & 1 \\ J_2 & J_2+1 & 1 \\ J & J & 1\end{matrix} \right\}
C^{J 0}_{(J_1+1)(0)(J_2+1)(0)}
\right]
\end{split}\end{equation*}

\begin{equation*}\begin{split}
&C^{(1,0,-1)}_{J_1 M_1 J_2 M_2 J M} =
\sqrt{\frac{3}{2 \pi}} (2 J_1+1) C^{J M}_{J_1 M_1 J_2 M_2} \cdot \\
&\left[
\sqrt{(J_2)(2 J_2-1)}
\left\{ \begin{matrix}J_1 & J_1 & 1 \\ J_2 & J_2-1 & 1 \\ J & J+1 & 1\end{matrix} \right\}
C^{(J+1) 0}_{(J_1)(0)(J_2-1)(0)} \sqrt{\frac{J}{2J+1}} \right. \\
&\left.
-\sqrt{(J_2+1)(2 J_2+3)}~
\left\{ \begin{matrix}J_1 & J_1 & 1 \\ J_2 & J_2+1 & 1 \\ J & J+1 & 1\end{matrix} \right\}
C^{(J+1) 0}_{(J_1)(0)(J_2+1)(0)} \sqrt{\frac{J}{2J+1}} \right. \\
&\left.
+\sqrt{(J_2)(2 J_2-1)}
\left\{ \begin{matrix}J_1 & J_1 & 1 \\ J_2 & J_2-1 & 1 \\ J & J-1 & 1\end{matrix} \right\}
C^{(J-1) 0}_{(J_1)(0)(J_2-1)(0)} \sqrt{\frac{J+1}{2J+1}} \right. \\
&\left.
-\sqrt{(J_2+1)(2 J_2+3)}~
\left\{ \begin{matrix}J_1 & J_1 & 1 \\ J_2 & J_2+1 & 1 \\ J & J-1 & 1\end{matrix} \right\}
C^{(J-1) 0}_{(J_1)(0)(J_2+1)(0)} \sqrt{\frac{J+1}{2J+1}}
\right]
\end{split}\end{equation*}

\begin{equation*}
\begin{split}
&C^{(0,0,-1)}_{J_1 M_1 J_2 M_2 J M} =
\sqrt{\frac{3}{2 \pi}} (2 J_1+1) C^{J M}_{J_1 M_1 J_2 M_2}  \cdot \\
&\left[
\sqrt{(J_2)(2 J_2-1)}
\left\{ \begin{matrix} J_1 & J_1   & 1 \\ J_2 & J_2-1 & 1 \\ J & J & 1 \end{matrix} \right\}
C^{J 0}_{(J_1)(0)(J_2-1)(0)} \right.
\left.
-\sqrt{(J_2+1)(2 J_2+3)}
\left\{ \begin{matrix} J_1 & J_1 & 1 \\ J_2 & J_2+1 & 1 \\ J & J & 1 \end{matrix} \right\}
C^{J 0}_{(J_1)(0)(J_2+1)(0)}
\right]
\end{split}
\end{equation*}

\begin{equation*}\begin{split}
&C^{(1,1,-1)}_{J_1 M_1 J_2 M_2 J M} =
\sqrt{\frac{3}{2 \pi}} C^{J M}_{J_1 M_1 J_2 M_2} \cdot \\
&\left[ 
\sqrt{(J_1+1)(J_2)(2 J_1-1)(2 J_2-1)}
\left\{ \begin{matrix}J_1 & J_1-1 & 1 \\ J_2 & J_2-1 & 1 \\ J & J+1 & 1\end{matrix} \right\}
C^{(J+1) 0}_{(J_1-1)(0)(J_2-1)(0)} \sqrt{\frac{J}{2J+1}} \right. \\
&\left.
- \sqrt{(J_1+1)(J_2+1)(2 J_1-1)(2 J_2+3)}
\left\{ \begin{matrix}J_1 & J_1-1 & 1 \\ J_2 & J_2+1 & 1 \\ J & J+1 & 1\end{matrix} \right\}
C^{(J+1) 0}_{(J_1-1)(0)(J_2+1)(0)} \sqrt{\frac{J}{2J+1}} \right. \\
&\left.
+ \sqrt{(J_1)(J_2)(2 J_1+3)(2 J_2-1)}
\left\{ \begin{matrix}J_1 & J_1+1 & 1 \\ J_2 & J_2-1 & 1 \\ J & J+1 & 1\end{matrix} \right\}
C^{(J+1) 0}_{(J_1+1)(0)(J_2-1)(0)} \sqrt{\frac{J}{2J+1}} \right. \\
&\left.
- \sqrt{(J_1)(J_2+1)(2 J_1+3)(2 J_2+3)}
\left\{ \begin{matrix}J_1 & J_1+1 & 1 \\ J_2 & J_2+1 & 1 \\ J & J+1 & 1\end{matrix} \right\}
C^{(J+1) 0}_{(J_1+1)(0)(J_2+1)(0)} \sqrt{\frac{J}{2J+1}} \right. \\
&\left.
+ \sqrt{(J_1+1)(J_2)(2 J_1-1)(2 J_2-1)}
\left\{ \begin{matrix}J_1 & J_1-1 & 1 \\ J_2 & J_2-1 & 1 \\ J & J-1 & 1\end{matrix} \right\}
C^{(J-1) 0}_{(J_1-1)(0)(J_2-1)(0)} \sqrt{\frac{J+1}{2J+1}} \right. \\
&\left.
- \sqrt{(J_1+1)(J_2+1)(2 J_1-1)(2 J_2+3)}
\left\{ \begin{matrix}J_1 & J_1-1 & 1 \\ J_2 & J_2+1 & 1 \\ J & J-1 & 1\end{matrix} \right\}
C^{(J-1) 0}_{(J_1-1)(0)(J_2+1)(0)} \sqrt{\frac{J+1}{2J+1}} \right. \\
&\left.
+ \sqrt{(J_1)(J_2)(2 J_1+3)(2 J_2-1)}
\left\{ \begin{matrix}J_1 & J_1+1 & 1 \\ J_2 & J_2-1 & 1 \\ J & J-1 & 1\end{matrix} \right\}
C^{(J-1) 0}_{(J_1+1)(0)(J_2-1)(0)} \sqrt{\frac{J+1}{2J+1}} \right. \\
&\left.
- \sqrt{(J_1)(J_2+1)(2 J_1+3)(2 J_2+3)}
\left\{ \begin{matrix}J_1 & J_1+1 & 1 \\ J_2 & J_2+1 & 1 \\ J & J-1 & 1\end{matrix} \right\}
C^{(J-1) 0}_{(J_1+1)(0)(J_2+1)(0)} \sqrt{\frac{J+1}{2J+1}}
\right]
\end{split}\end{equation*}

where $C_{{J_1}{M_1}{J_2}{M_2}}^{JM}$ are the Clebsch-Gordan coefficients  [Chapter 8 in Ref. \onlinecite{Varshalovich}], and the quantities in braces are Wigner $6j$ and $9j$ symbols [Chapters 9 and 10 in Ref. \onlinecite{Varshalovich}].

The coefficients $\left\{ {{u''}_{mn}^{(2\omega )},{v''}_{mn}^{(2\omega )}} \right\}$ for both the $\gamma$ bulk and the surface normal polarization source can be expressed as:
\begin{equation*}
\begin{split}
&{u''}_{mn}^{(2\omega )} =
\left( - \frac{a}{4} \right) i \sqrt {n(n + 1)} \frac{g_{mn}({k_i}(\omega )R)}{{k_0}(\omega )R}
+ \left( - \frac{d}{8} \right) \frac{{{\varepsilon _0}}}{{{\varepsilon _i}(2\omega )}}i\sqrt {n(n + 1)} \frac{{{f_{mn}}({k_i}(\omega )R)}+{{g_{mn}}({k_i}(\omega )R)}}{{{k_0}(\omega )}R} \\
&{v''}_{mn}^{(2\omega )} = 0
\end{split}
\end{equation*}
where

\begin{equation*}
\begin{split}
&g_{mn}({k_i}(\omega )R) = \sum\limits_{{n_1}{m_1}}{} \sum\limits_{{n_2}{m_2}} {} \\
&\left\{
A_{{m_1}{n_1}}^{( - 1)}A_{{m_2}{n_2}}^{( - 1)}
\left[
{\sqrt {\frac{{{n_1}}}{{2{n_1} + 1}}} \sqrt {\frac{{{n_2}}}{{2{n_2} + 1}}} W_{nm}^{{n_1} - 1,{n_1},{m_1},{n_2} - 1,{n_2},{m_2}} +}
\right.
\left.
{\sqrt {\frac{{{n_1} + 1}}{{2{n_1} + 1}}} \sqrt {\frac{{{n_2} + 1}}{{2{n_2} + 1}}} W_{nm}^{{n_1} + 1,{n_1},{m_1},{n_2} + 1,{n_2},{m_2}}}
\right] \right. \\
&\left.
- A_{{m_1}{n_1}}^{( - 1)}A_{{m_2}{n_2}}^{( - 1)}
\left[
{\sqrt {\frac{{{n_1}}}{{2{n_1} + 1}}} \sqrt {\frac{{{n_2} + 1}}{{2{n_2} + 1}}} W_{nm}^{{n_1} - 1,{n_1},{m_1},{n_2} + 1,{n_2},{m_2}} +}
\right.
\left.
{\sqrt {\frac{{{n_1} + 1}}{{2{n_1} + 1}}} \sqrt {\frac{{{n_2}}}{{2{n_2} + 1}}} W_{nm}^{{n_1} + 1,{n_1},{m_1},{n_2} - 1,{n_2},{m_2}}}
\right]
\right\}
\end{split}
\end{equation*}

\begin{align*}
&f_{mn}({k_i}(\omega )R) = \sum\limits_{{n_1},{m_1}}{} \sum\limits_{{n_2},{m_2}}{} \\
&\left\{
A_{{m_1}{n_1}}^{(1)}A_{{m_2}{n_2}}^{(1)}
\left[
{\sqrt {\frac{{{n_1} + 1}}{{2{n_1} + 1}}} \sqrt {\frac{{{n_2} + 1}}{{2{n_2} + 1}}} W_{nm}^{{n_1} - 1,{n_1},{m_1},{n_2} - 1,{n_2},{m_2}} +}
\right.
\left.
{\sqrt {\frac{{{n_1}}}{{2{n_1} + 1}}} \sqrt {\frac{{{n_2}}}{{2{n_2} + 1}}} W_{nm}^{{n_1} + 1,{n_1},{m_1},{n_2} + 1,{n_2},{m_2}}}
\right]
\right. \\
&\left.
+ A_{{m_1}{n_1}}^{(1)}A_{{m_2}{n_2}}^{(1)}
\left[
{\sqrt {\frac{{{n_1} + 1}}{{2{n_1} + 1}}} \sqrt {\frac{{{n_2}}}{{2{n_2} + 1}}} W_{nm}^{{n_1} - 1,{n_1},{m_1},{n_2} + 1,{n_2},{m_2}} +}
\right.
\left.
{\sqrt {\frac{{{n_1}}}{{2{n_1} + 1}}} \sqrt {\frac{{{n_2} + 1}}{{2{n_2} + 1}}} W_{nm}^{{n_1} + 1,{n_1},{m_1},{n_2} - 1,{n_2},{m_2}}  }
\right] \right.\\
&\left.
+ A_{{m_1}{n_1}}^{(0)}A_{{m_2}{n_2}}^{(0)}~~W_{nm}^{{n_1},{n_1},{m_1},{n_2},{n_2},{m_2}} \right.\\
&\left.
%+ A_{{m_1}{n_1}}^{( - 1)}A_{{m_2}{n_2}}^{( - 1)}
%\left[
%{\sqrt {\frac{{{n_1}}}{{2{n_1} + 1}}} \sqrt {\frac{{{n_2}}}{{2{n_2} + 1}}} W_{nm}^{{n_1} - 1,{n_1},{m_1},{n_2} - 1,{n_2},{m_2}}       +}
%\right.
%\left.
%{\sqrt {\frac{{{n_1} + 1}}{{2{n_1} + 1}}} \sqrt {\frac{{{n_2} + 1}}{{2{n_2} + 1}}} W_{nm}^{{n_1} + 1,{n_1},{m_1},{n_2} + 1,{n_2},{m_2}}}
%\right] \right. \\
%&\left.
%- A_{{m_1}{n_1}}^{( - 1)}A_{{m_2}{n_2}}^{( - 1)}
%\left[
%{\sqrt {\frac{{{n_1}}}{{2{n_1} + 1}}} \sqrt {\frac{{{n_2} + 1}}{{2{n_2} + 1}}} W_{nm}^{{n_1} - 1,{n_1},{m_1},{n_2} + 1,{n_2},{m_2}} +}
%\right.
%\left.
%{\sqrt {\frac{{{n_1} + 1}}{{2{n_1} + 1}}} \sqrt {\frac{{{n_2}}}{{2{n_2} + 1}}} W_{nm}^{{n_1} + 1,{n_1},{m_1},{n_2} - 1,{n_2},{m_2}}  }
%\right] \right. \\
%&\left.
+ A_{{m_1}{n_1}}^{(1)}A_{{m_2}{n_2}}^{(0)}
\left[
{\sqrt {\frac{{{n_1} + 1}}{{2{n_1} + 1}}} W_{nm}^{{n_1} - 1,{n_1},{m_1},{n_2},{n_2},{m_2}} +}
\right.
\left.
{\sqrt {\frac{{{n_1}}}{{2{n_1} + 1}}} W_{nm}^{{n_1} + 1,{n_1},{m_1},{n_2},{n_2},{m_2}}      }
\right] \right. \\
&\left.
+ A_{{m_1}{n_1}}^{(0)}A_{{m_2}{n_2}}^{(1)}
\left[
{\sqrt {\frac{{{n_2} + 1}}{{2{n_2} + 1}}} W_{nm}^{{n_1},{n_1},{m_1},{n_2} - 1,{n_2},{m_2}} +}
\right.
\left.
{\sqrt {\frac{{{n_2}}}{{2{n_2} + 1}}} W_{nm}^{{n_1},{n_1},{m_1},{n_2} + 1,{n_2},{m_2}}      }
\right]
\right\}
\end{align*}

\begin{align*}
W_{LM}^{L1,J1,M1,L2,J2,M2} =
{( - 1)^{{J_2} + {L_1} + L}}
\sqrt {\frac{{(2{J_1} + 1)(2{J_2} + 1)(2{L_1} + 1)(2{L_2} + 1)}}{{4\pi (2L + 1)}}}
\left\{ {\begin{array}{*{20}{c}}
{{L_1}}&{{L_2}}&L\\
{{J_2}}&{{J_1}}&1
\end{array}} \right\}
C_{{L_1}0{L_2}0}^{L0}~C_{{J_1}{M_1}{J_2}{M_2}}^{LM}
\end{align*}

\end{widetext}
%\twocolumngrid

\newpage

\onecolumngrid
\begin{acknowledgments}
This work was partly supported by the Italian Miur through the project PON01-02782. L.D.N. acknowledges the support of the NSF Career Award No. ECCS-0846651.
\end{acknowledgments}
\twocolumngrid

% Create the reference section using BibTeX:
\bibliography{manuscript}

\end{document}